\documentclass[%
 aip,
 amsmath,amssymb,
 reprint,%
]{revtex4-1}

\usepackage{graphicx}
\usepackage{subcaption}
\usepackage{dcolumn}
\usepackage{bm}
\usepackage{comment}

\usepackage[utf8]{inputenc}
\usepackage[T1]{fontenc}
\usepackage{mathptmx}
\usepackage{etoolbox}

\makeatletter
\def\@email#1#2{%
 \endgroup
 \patchcmd{\titleblock@produce}
  {\frontmatter@RRAPformat}
  {\frontmatter@RRAPformat{\produce@RRAP{*#1\href{mailto:#2}{#2}}}\frontmatter@RRAPformat}
  {}{}
}%
\makeatother
\begin{document}

\preprint{AIP/123-QED}

\title[Investigating the Lower Hybrid Drift Instability in Reconnecting Current Sheets Using a Hybrid Kinetic Model (ssV Code)]{Investigating the Lower Hybrid Drift Instability in Reconnecting Current Sheets Using a Hybrid Kinetic Model (ssV Code)}
\author{S. Thatikonda}
\email[Corresponding author: ]{sreenivasa.thatikonda@ipp.mpg.de} 
\affiliation{Max-Planck Institute for Plasma Physics, Boltzmannstrasse 2, Garching, 85748, Bavaria, Germany}

\author{F. N. De Oliveira-Lopes}
\affiliation{Max-Planck Institute for Plasma Physics, Boltzmannstrasse 2, Garching, 85748, Bavaria, Germany}
\affiliation{Centre for mathematical Plasma Astrophysics, KU Leuven, Celestijnenlaan 200B, Leuven, Belgium}

\author{A. Mustonen}
\affiliation{Max-Planck Institute for Plasma Physics, Boltzmannstrasse 2, Garching, 85748, Bavaria, Germany}
\author{K. Pommois}
\affiliation{Max-Planck Institute for Plasma Physics, Boltzmannstrasse 2, Garching, 85748, Bavaria, Germany}
\author{D. Told}
\affiliation{Max-Planck Institute for Plasma Physics, Boltzmannstrasse 2, Garching, 85748, Bavaria, Germany}
\author{F. Jenko}
\affiliation{Max-Planck Institute for Plasma Physics, Boltzmannstrasse 2, Garching, 85748, Bavaria, Germany}

\date{\today}

\begin{abstract}
We investigate the nonlinear evolution of the lower hybrid drift instability (LHDI) in reconnecting current sheets using a hybrid kinetic simulation model implemented in the Super Simple Vlasov (ssV) code. The model treats ions kinetically and electrons with a drift-kinetic approximation, solving self-consistent coupled electrostatic and electromagnetic fields. A parametric study explores the effects of mass ratio, temperature ratio, plasma beta, and sheet thickness. In electrostatic cases, LHDI remains localized at the sheet edges, flattening density gradients. In electromagnetic regimes, turbulence induced by LHDI generates magnetic perturbations that kink the current sheet and enhance anomalous resistivity. These dynamics may facilitate fast magnetic reconnection under certain conditions. Our results bridge prior theoretical predictions and simulations, emphasizing the importance of kinetic instabilities in reconnection physics.
\end{abstract}

\maketitle
\section{Introduction}
 Magnetic reconnection in collisionless plasmas is widely believed to be enabled by small-scale “anomalous” resistivity, which breaks the frozen-in field condition via turbulence in the current sheet~\cite{daughton2003, daughton2004, hesse2019, yamada2010, zweibel2009, munoz2015}. Various microscopic instabilities can generate the necessary fluctuations, including the tearing mode, modified two-stream instability, drift-kink/sausage modes, and the lower hybrid drift instability (LHDI)~\cite{coppi1966, pritchett2001, daughton2003, daughton2004, lin2008, yoon2008, innocenti2016, rahman2021, munoz2015, schmitz2006}. 
 
 In particular, LHDI has long been considered a prime candidate for producing the turbulent resistivity required for fast reconnection~\cite{huba1977, daughton2004, yoon2008, lin2008, innocenti2016, rahman2021, treumann2001}. The LHDI is a cross-field instability driven by density gradients in the presence of a sheared plasma flow, with a characteristic frequency near the lower hybrid frequency~\cite{huba1977, yoon2008}. Early analyses, however, pointed out a potential limitation, that in thin current sheets, the fastest-growing LHDI modes are localized to the sheet edges and do not strongly perturb the central region~\cite{daughton2003, daughton2004, lin2008}. Because LHDI modes predominantly develop at the edges of thin current sheets, early theoretical studies raised doubts about whether they could effectively enhance resistivity at the X-line where magnetic reconnection initiates~\cite{daughton2003, yoon2008}. 
 
 This skepticism persisted in the community for some time~\cite{treumann2001}. Observational studies at the magnetopause, including recent MMS mission data, have confirmed that LHDI-induced fluctuations are predominantly localized at current sheet boundaries, with negligible impact on the core current, supporting earlier theoretical predictions~\cite{graham2019, phan2018, ergun2020, stawarz2021}.
 
 More recent theoretical and simulation work has revisited this issue, revealing a more complex picture. Linear kinetic theory and gyrokinetic models have demonstrated that LHDI modes can exist over a broad range of wavelengths, from electron gyroradius scales up to near ion scales~\cite{yoon2008, lin2008, rahman2021, gary1993}, and that the presence of a finite magnetic guide field or finite parallel wave number can significantly alter the instability’s structure~\cite{lin2008, innocenti2016, ng2023}.
 
 Notably, Daughton et al.~\cite{daughton2004} performed fully kinetic particle-in-cell simulations with realistic mass ratio and found that the nonlinear LHDI can significantly modify the current sheet. Although the LHDI initially develops at the sheet edges, its nonlinear growth drives an enhanced electron flow in the central region, leading to a strong bifurcation (splitting) of the current layer and substantial anisotropic heating of electrons~\cite{daughton2004, yoon2008, innocenti2016, ng2023}. This turbulence-induced restructuring was found to dramatically enhance the collisionless tearing mode, potentially triggering rapid magnetic reconnection once the current sheet thins to kinetic scales~\cite{daughton2004, yoon2008, innocenti2016, ng2023}. These findings indicate that LHDI, previously thought to be benign for reconnection, may in fact play a crucial role under the right conditions~\cite{daughton2004, innocenti2016, ng2023, yoon2008}. 
 
 Despite these advances, many open questions remain regarding LHDI’s exact role in reconnection dynamics. Does the LHDI-triggered turbulence cause fast reconnection, or is it merely a byproduct of an already reconnecting current sheet~\cite{daughton2004, ng2023} How do key plasma parameters (mass ratio, temperature ratio, plasma beta, current sheet thickness) influence the instability’s behavior in the fully nonlinear regime~\cite{yoon2008, lin2008, innocenti2016} Addressing these questions requires simulations that span the disparate electron-ion scales at an affordable computational cost~\cite{told2011, pueschel2011, munoz2015}. 
 
 In this work, we employ a hybrid gyrokinetic approach using the recently developed ssV code~\cite{your2025paper} to investigate the nonlinear LHDI in reconnecting current sheets. By treating ions kinetically and electrons with a drift kinetic model, we capture essential kinetic physics (e.g. LHDI, Landau resonance, wave-particle scattering) while mitigating the scale separation challenge of full kinetic simulations. 
 
 This study extends previous fully kinetic simulations and reduced fluid models by systematically scanning a range of mass ratios (including values approaching realistic), temperature ratios, plasma $\beta_e$, and current sheet thicknesses~\cite{daughton2004, yoon2008, lin2008, ng2023, rahman2021}. We separately examine purely electrostatic and fully electromagnetic cases to isolate how magnetic fluctuations alter the instability and its impact on reconnection~\cite{lin2008, innocenti2016}. Through this comprehensive parametric study, we aim to clarify the conditions under which LHDI provides significant anomalous resistivity or viscosity to the current sheet, and whether those effects can initiate or enhance collisionless reconnection. The results have implications for understanding magnetospheric substorm onset, laboratory reconnection experiments, and the general interplay between microturbulence and magnetic reconnection in plasmas~\cite{hesse2019, yamada2010, treumann2001}.

The remainder of this paper is organized as follows. In Sec.~\ref{sec:theory}, we outline the theoretical background of the lower hybrid drift instability, highlighting its linear and nonlinear characteristics relevant to reconnection. Sec.~\ref{sec:setup} describes the hybrid kinetic model and simulation setup used in the \textit{ssV} code, including the parameter space explored. In Sec.~\ref{sec:results}, we present simulation results for the electrostatic and electromagnetic cases, respectively, including growth rate trends, nonlinear mode structures, and heating diagnostics. The role of anomalous resistivity and viscosity is discussed in Sec.~\ref{sec:discussion}, supported by both theoretical arguments and simulation evidence. Finally, Sec.~\ref{sec:conclusion} summarizes our key findings and discusses implications for both space and laboratory plasmas.

\section{Theoretical Background}\label{sec:theory}
The lower hybrid drift instability is an electrostatic-dominant instability that arises in magnetized current sheets due to the diamagnetic drift of plasma in the presence of sharp density gradients~\cite{gary1993,daughton2003}. In a Harris-type current sheet (with antiparallel magnetic fields and a strong density gradient across the sheet)~\cite{Ricci2005}, ions and electrons experience different drift velocities, which can drive waves at the lower hybrid frequency $\omega_{LH} = \sqrt{\Omega_{ci}\Omega_{ce}}$ (between the ion and electron gyrofrequencies). 

For a simplified case with no guide field, linear theory predicts two primary branches of the LHDI~\cite{daughton2003,yoon2008}. The fast electron-scale branch (often termed mode A) is predominantly electrostatic and localized at the sheet edges~\cite{daughton2004,Ricci2005}, where the density gradient is largest. This mode has a characteristic perpendicular wavenumber $k_\perp \rho_e \sim 1$ (on the order of the electron gyroradius scale) and a growth rate on the order of $\omega_{LH}$. Because it is concentrated in the low-density “flanks” of the current sheet, mode A typically produces fluctuations in density and electric field there, with relatively small perturbations to the magnetic field or central current. A slower ion-scale branch of LHDI (mode B) can develop when the current sheet is sufficiently thin (on the order of an ion gyroradius or smaller)~\cite{daughton2004,yoon2008}. This branch peaks near the sheet center and involves a significant electromagnetic component. It has $k_\perp$ on the scale of the geometric mean of ion and electron gyroradii (e.g. $k_\perp \sqrt{\rho_i\rho_e} \sim 1$) and a growth rate comparable to the ion cyclotron frequency $\Omega_{ci}$, which is lower than that of mode A. Mode B arises from the coupling of the lower hybrid drift waves with electromagnetic perturbations of the current sheet; physically, it can be viewed as the LHDI driving a ripple or oscillation of the current layer itself. 

Importantly, this mode can reside in the central region of the sheet (where the current is maximum), and thus it has the potential to directly influence the reconnection site by modulating the current density and inducing electric fields there. Indeed, earlier kinetic simulations reported the appearance of longer-wavelength electromagnetic fluctuations in the sheet core after the saturation of the fastest LHDI modes~\cite{innocenti2016,rahman2021}. These were sometimes interpreted as separate instabilities like the drift-kink mode, but are now understood to be an extension of the LHDI spectrum (an electromagnetic LHDI branch) excited by nonlinear mode coupling. 

At even longer scales, on the order of the ion inertial length $d_i$, the current sheet can also undergo kinking or twisting instabilities (sometimes called the ion-ion kink or Kelvin-Helmholtz type modes)~\cite{daughton2004,Ricci2005}. This mode C in the above classification has $k_\perp d_i \sim 0.5$–2 and growth rates typically a small fraction of $\Omega_{ci}$. It corresponds to a large-scale deformation of the entire current channel. While the kink mode appears structurally similar to its MHD counterpart, in collisionless plasmas it is triggered by kinetic mechanisms rather than collisional MHD processes. While mode C is distinct from LHDI in origin, it can be triggered or facilitated by the nonlinear evolution of LHDI: the turbulence from modes A and B can create seed perturbations and modify the current profile in ways that lower the threshold for the kink mode~\cite{daughton2004,Ricci2005}. For example, simulations have shown that after LHDI saturates, the current sheet may become susceptible to a secondary kink or flapping motion. In a fully three-dimensional system, all three types of modes (A: electrostatic LHDI, B: electromagnetic LHDI, C: kink-type modes) may coexist or occur sequentially, influencing each other’s growth. 

The nature of LHDI can also be altered by the presence of a guide magnetic field (a component of magnetic field along the current direction). A finite guide field introduces a parallel (to the magnetic field) wave number $k_{\parallel}$ for perturbations and can make the instability more three-dimensional~\cite{lin2008,yoon2008}. Two-fluid theory and kinetic studies indicate that with a guide field, the most unstable LHDI modes become oblique (propagating with $k_{\parallel}\neq0$)~\cite{lin2008,yoon2008}. The guide field tends to reduce the growth of the edge mode A, while enabling modes that involve parallel electron motion. In such cases, the LHDI fluctuations can acquire a substantial parallel electric field component and resonate with electrons moving along the field line, leading to enhanced electron scattering and anomalous resistivity~\cite{innocenti2016,rahman2021}. Moreover, particle simulations under finite guide field have identified multiple LHDI branches analogous to modes A, B, C discussed above, but with different polarization~\cite{lin2008,yoon2008}: for instance, one mode dominated by electrostatic $E_z$ and $B_y$ perturbations, and another mode where electromagnetic perturbations ($B_x$) are equally important. As the guide field strengthens, the central (mode B-like) instability can become dominant, and its character transitions toward that of a drift-kink mode, which produces bending of magnetic field lines along the out-of-plane direction (\(\hat{x}\)), perpendicular to the simulation plane defined by \(y\) and \(z\). 

In summary, the LHDI in a current sheet can exhibit both electrostatic and electromagnetic behavior, with the balance between them controlled by parameters such as sheet thickness, plasma beta (which sets the importance of magnetic pressure), and any guide field. The impact of LHDI on magnetic reconnection has been a subject of active research and debate~\cite{daughton2003,Ricci2005,innocenti2016}. On one hand, LHDI-driven fluctuations can produce anomalous resistivity, effectively mimicking collisions by scattering electrons and impeding the current flow. This can break the frozen-in condition at the reconnection site and potentially speed up the reconnection rate. On the other hand, if the instability is confined to the sheet edges, it may primarily cause anomalous viscosity or momentum transport – for example, by transferring momentum from the drifting ions to the background, thus altering the velocity shear (which is more related to the drift-kink mode) rather than directly breaking field lines. Early work concluded that edge-localized LHDI modes could not directly enhance reconnection because they fail to produce turbulence in the X-line region where it’s needed. 

However, more recent nonlinear simulations suggest a more nuanced scenario: the LHDI can indirectly trigger reconnection by thinning the current sheet and forming secondary instabilities. As LHDI saturates, it often flattens the density profile at the sheet edges, reducing pressure there and causing a net inward pressure force. This can lead to a collapse (thinning) of the current sheet, bringing oppositely directed field lines closer together and amplifying the central current density~\cite{Ricci2005,rahman2021}. In a sheet near the marginal stability threshold, this thinning can precipitate the onset of the collisionless tearing mode (the direct driver of reconnection) much faster than would occur without LHDI. 

Conversely, in thicker or higher-\(\beta_e\) sheets, LHDI-induced transport may weakly smooth out edge gradients, exerting a mild stabilizing effect on tearing modes, but without significantly altering the overall reconnection rate. Whether LHDI triggers reconnection or just accompanies it thus depends on the plasma parameters. It is this interplay—between anomalous resistivity vs. viscosity, and between LHDI-driven turbulence vs. tearing-driven dynamics—that we seek to clarify with our simulations.

\section{Numerical Model and Setup}\label{sec:setup}
To capture the multi-scale physics of LHDI in a reconnecting current sheet, we utilize the hybrid kinetic simulation code ssV. The code name ssV stands for “Super Simple Vlasov,” reflecting its design as a semi-Lagrangian Vlasov solver~\cite{your2025paper}. In this hybrid model, ions are treated with full kinetic physics (i.e. their full Vlasov equation is solved without approximation), while electrons are treated with drift-kinetic physics~\cite{Ricci2005,ng2023}. The ion dynamics are governed by the full Vlasov equation:
\begin{equation}
\frac{\partial f_i}{\partial t}
+ \mathbf{v} \cdot \nabla f_i
+ \frac{q_i}{m_i} \left( \mathbf{E} + \mathbf{v} \times \mathbf{B} \right) \cdot \nabla_{\mathbf{v}} f_i = 0
\end{equation}
whereas the drift-kinetic equation for electrons, appropriate in the limit $\omega \ll \Omega_{ce}$, takes the form:
\begin{align}
\frac{\partial f_e}{\partial t}
&+ \left( v_{gy,\parallel} + \frac{e}{m} A_{1\parallel} \right) \cdot \nabla_{gy} f_e \nonumber \\
&+ \frac{\hat{b}}{eB_0} \times 
\left( e \nabla \phi_1 - e v_{gy,\parallel} \nabla A_{1\parallel} \right) \cdot \nabla_{gy} f_e \nonumber \\
&- \left( \frac{e}{m} \nabla \phi_1 - \frac{e}{m} v_{gy,\parallel} \nabla A_{1\parallel} \right) 
\cdot \frac{\partial f_e}{\partial v_{gy,\parallel}} = 0
\end{align}
Quasi-neutrality is imposed via:
\begin{equation}
n_i = n_e
\end{equation}
with $n_i = \int f_i \, d^3v$ for ions and $n_e = \int f_e \, dv_\parallel$.

The self-consistent fields are obtained by solving coupled Poisson and Ampère equations. The generalized Poisson equation is:
\begin{align}\label{eq:poissoneq}
\frac{1}{4\pi}\nabla_{\perp}^{2}\phi_{1}(x)\left(4\pi\frac{\rho_{th}^{2}}{\lambda_{D}^{2}} - 1\right)\nonumber\\
&  + u_{e \parallel}(x) \frac{\rho_{th}^{2}}{\lambda_{D}^{2}} \nabla_{\perp}^2 A_{1 \parallel}(\mathbf{x})\nonumber \\
&  = \sum_i q_i n_i(\mathbf{x}) + e n_e(\mathbf{x})
\end{align}

where $\rho_{th}$ is the ion thermal Larmor radius, $\lambda_D$ is the Debye length, and $u_{e\parallel}$ is the parallel electron flow velocity, the factor $\left(\tfrac{4\pi\rho_{\mathrm{th}}^{2}}{\lambda_{D}^{2}} - 1\right)$ 
appearing in front of $\nabla_{\perp}^{2}\phi_{1}$ represents 
the effective dielectric coefficient arising from gyrokinetic electron polarization.

The parallel component of the vector potential $A_{1\parallel}$ is advanced using the coupled Ampère equation:
\begin{align}\label{eq:Ampereeq}
\frac{c}{4\pi}\nabla_{\perp}^{2} A_{1\parallel}\left(1+\frac{\beta_e}{2}\right)\nonumber\\
&  + u_{e \parallel}\frac{\rho_e^2}{\lambda_D^2}\nabla_{\perp}^2\phi_1(\textbf{x})\nonumber\\
& =\frac{e^2}{m_e c} n_e A_{1 \parallel}(\textbf{x})- I_e+\sum_i I_{i \parallel}
\end{align}  

where $\rho_e$ is the electron Larmor radius, $\beta_e$ is the electron beta, and $I_{e}, I_{i\parallel}$ are the electron and ion parallel currents. The factor $(1+\beta_{e}/2)$ represents the electron pressure contribution to the magnetization current, while the coupling term proportional to $\nabla_{\perp}^{2}\phi_{1}$ arises from the electron parallel flow.

The drift-kinetic electron model averages out the fast gyromotion of electrons (assuming $\omega \ll \Omega_{ce}$), which is a valid approximation in the lower-hybrid frequency range of interest~\cite{gary1993,rahman2021}. This approach dramatically reduces the computational cost associated with following electron cyclotron motion, yet retains essential electron kinetic effects such as Landau damping and wave-particle resonance. The ions, on the other hand, are kept fully kinetic to correctly resolve ion gyro-radius scale physics (critical for LHDI and ion-cyclotron-range instabilities). 

The ions and electrons are coupled through Maxwell’s equations in the electrostatic limit (Poisson’s equation for the electric potential $\phi$) and through a coupled Poisson and Ampère’s law for the magnetic field perturbations. In the ssV code, we solve the quasineutrality condition $\nabla \cdot \mathbf{E}=0$ (appropriate for low-frequency, sub-light-speed phenomena) to obtain the electrostatic potential, and we advance the parallel component of Ampère’s law to compute the evolution of the magnetic field (in the presence of current perturbations). This permits electromagnetic fluctuations ($\delta B$) to develop self-consistently in the simulations when desired, while avoiding the numerical stiffness of full Maxwell-Faraday coupling (we neglect displacement current since $v \ll c$ in our regime). Care is taken to properly resolve the so-called “Ampère cancellation problem” that can arise in gyrokinetic simulations due to the delicate balance of plasma currents~\cite{Mandell2020,Pezzi2019,Dannert2004}. 

The ssV code uses high-order semi-Lagrangian schemes for the Vlasov equation (including SLMP~\cite{Tanaka2017}, a fifth-order flux-conservative scheme and a specialized monotonic limiter), ensuring low numerical dissipation and accurate resolution of fine velocity-space structures. 

All quantities in our simulations are expressed in normalized units. Lengths are normalized to the ion gyroradius $\rho_i$ (evaluated using a reference magnetic field $B_0$ and ion thermal speed), unless otherwise specified. Times are normalized to the inverse ion cyclotron frequency $\Omega_{ci}^{-1}$, and velocities to the ion thermal speed $v_{th,i}$ and the electron thermal speed $v_{th,e}$ respectively. The electric field is normalized by $B_0 v_{th,i}$ and magnetic field perturbations by $B_0$. Unless stated, the background magnetic field (in the inflow region far from the current sheet) is $B_0$ and points in $\pm \hat{x}$ (antiparallel configuration).  Both ions and electrons are initialized with Maxwellian velocity distributions with specified temperatures $T_i$ and $T_e$, and the ratio $T_i/T_e$ is an input parameter. 

We consider a 2D simulation domain in the $y$–$z$ plane: $z$ is the direction perpendicular to the current sheet (gradient direction), and $y$ is along the sheet (current flows in the $x$-direction out of the plane). The simulation domain size in $z$ is chosen to be several times the current sheet half-thickness so that plasma on both sides of the sheet is included; in $y$, we take a domain length sufficient to encompass one wavelength of the dominant instability mode (for LHDI, this is typically on the order of several ion gyroradii along the current direction). Periodic boundary conditions are applied in the $y$ and $z$ directions. These boundary conditions are appropriate for the Harris current sheet (effectively modeling a double Harris sheet, one at the mid-plane and one at the $z$-domain boundaries, to maintain periodicity). The initial equilibrium is a Harris-like current sheet~\cite{Ricci2005,lin2008}: 
\begin{equation}
    B_y(z) = B_0 \tanh(z/L)
\end{equation} 
with a peak current (carried by drift electrons) at $z=0$. Here $L$ is the current sheet half-thickness (the scale over which $B_y$ transitions from $+B_0$ to $-B_0$). The plasma density is highest at the sheet center and follows pressure balance:
\begin{equation}
    n(z) = n_0 sech^{2}(z/L) + n_{\mathrm{bg}}
\end{equation} 
where $n_{\mathrm{bg}}=0.2 n_0$ is a uniform background density far from the sheet and $n_0 = 1.0$. In most runs we take a low background density ($n_{\mathrm{bg}} \ll n_0$) so that the sheet edges have a strong density gradient (this favors LHDI). The plasma beta is defined as $\beta_e = 2\mu_0 (n_e T_e)/B_0^2$. We choose values of $\beta_e$ in the range 0.01–0.2 to explore regimes from strongly magnetized to moderately high-beta plasma.
\begin{figure}[h]
    \centering
    \includegraphics[width=0.5\textwidth]{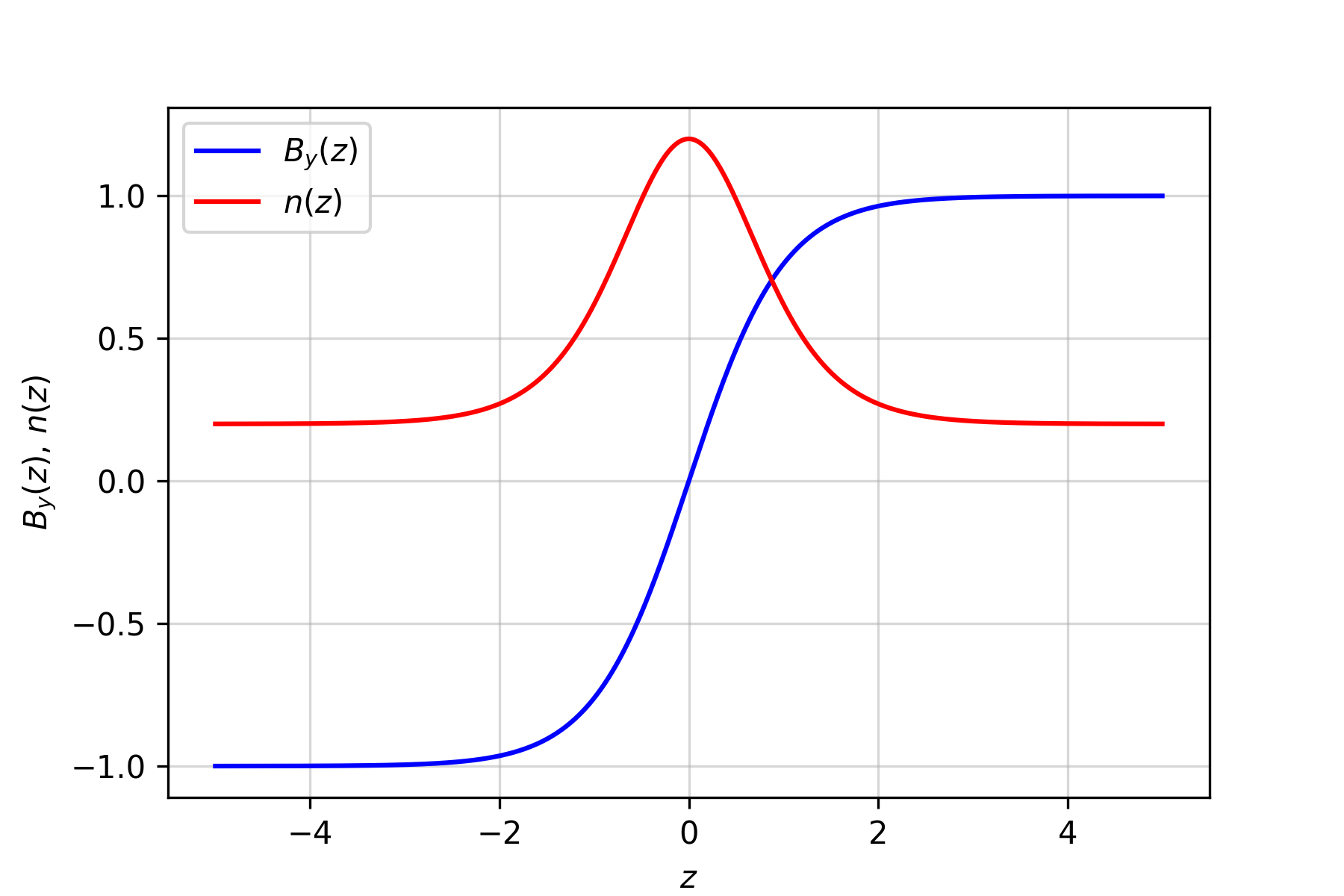}
    \caption{Initial Harris equilibrium profiles for the magnetic field 
    $B_y(z)$ and plasma density $n(z)$, with half-thickness $L = 1\,\rho_i$. 
    The density peaks at the current sheet center ($z = 0$), while the 
    magnetic field reverses direction across the sheet.}
    \label{fig:initial_profiles}
\end{figure}
Figure~\ref{fig:initial_profiles} shows the sample initial equilibrium profiles 
used in this study, with plasma density $n(z)$ and magnetic field $B_y(z)$ 
following the Harris current sheet configuration. The density peaks at the 
current sheet center ($z = 0$), while the magnetic field reverses across the 
sheet over a half-thickness $L = 1\,\rho_i$.
The ions are initialized with a fully kinetic Maxwellian distribution function,
\begin{equation}
f_i(z, \mathbf{v}) = \frac{n_{i}(z)}{(\pi )^{3/2}}
\exp\left[-(v_y - U_{i,y})^2 + v_x^2 + v_z^2\right]
\end{equation}
where the ion diamagnetic drift, arising from the density gradient across the sheet, is directed opposite to the electron drift that primarily supports the current.
\[
U_{i,y} = +\frac{T_i}{q_i B_0 L}
\]

Electrons are initialized with a drift-kinetic Maxwellian, depending only on the guiding-center parallel velocity $v_{\parallel}$ 
\begin{equation}
f_e(z, \mathbf{v}) = \frac{n_{e}(z)}{(\pi )^{1/2}}
\exp\left[-(v_{\parallel} - U_{e,\parallel})^2 \right]
\end{equation}
the electron drift velocity
\[
U_{e,\parallel} \simeq -\frac{T_e}{q_e B_0 L}
\]
To seed the LHDI, a small perturbation is applied to the ion distribution function in the $y$--$z$ plane. 
The perturbation is introduced through a stream function of the form:
\begin{equation}
\delta \psi(y, z) = \psi_0 \cos\left(\frac{2\pi y}{L_y}\right) \cos\left(\frac{2\pi z}{L_z}\right)
\end{equation}
with a small amplitude 
\[
\psi_0 = 0.001
\]
The perturbed ion distribution function is then expressed as
\begin{equation}
f_i^{\text{pert}}(y, z, \mathbf{v}) = f_i^0(y, z, \mathbf{v}) 
\left[1 + \delta \psi(y, z)\right]
\end{equation}
where $f_i^0$ is the equilibrium Harris distribution. This perturbation excites long-wavelength modes localized near the current sheet edges, allowing the LHDI to grow self-consistently.

\noindent\textbf{Parameter Scan and Cases:}
We conducted a systematic scan over key dimensionless parameters to study their effect on LHDI in the current sheet~\cite{daughton2003,Ricci2005,ng2023}:

Ion-to-electron mass ratio ($m_i/m_e$): Values of 5, 10, 25, 50, 100, 250 and 500 were used. Lower values are far from physical (real mass ratio $\sim1836$) but are computationally easier; by increasing up to 500 we approach more realistic scale separation. This allows us to assess how the instability scales toward a realistic mass ratio. All other parameters (temperature, density) were adjusted such that $\beta_e$ and other non-dimensional quantities remained the same across different $m_i/m_e$ cases.

Temperature ratio ($T_i/T_e$): Values of 1, 5, 10, 15, 20, 25 were examined. Keeping total $\beta_e$ fixed, varying $T_i/T_e$ effectively changes the partition of plasma pressure between ions and electrons. This influences the diamagnetic drift speeds and the relative response of ions vs. electrons to fields, hence affecting LHDI drive. High $T_i/T_e$ means hotter ions (and/or colder electrons), which tends to enhance LHDI drive because the ion diamagnetic drift is larger relative to electron thermal motion.

Plasma beta ($\beta_e$): We sampled $\beta_e=0.01, 0.05, 0.10, 0.15, 0.20$. Lower beta means a stronger magnetic field for a given pressure, which generally leads to higher lower-hybrid frequency and more electrostatic behavior (since $\omega_{LH}$ is higher and $\Omega_{ci}$ dominates). Higher beta means a weaker field relative to pressure, possibly allowing greater magnetic perturbations (since the field is more easily perturbed). This parameter thus controls the transition from electrostatic to electromagnetic regimes of LHDI.

Current sheet thickness ($2L$): We use half-thickness $L$ (the scale length of the sheet) values of $0.5 \rho_i, 0.75 \rho_i, 1 \rho_i,$ and $ 1.5 \rho_i$. In terms of full thickness (distance between asymptotic $B_0$), this is about $1.0 \rho_i$ up to $3.0 \rho_i$. Thin sheets ($L < \rho_i$) are near or below the ion gyroradius scale and are expected to strongly excite LHDI (including the electromagnetic branch). Thicker sheets should be more stable to LHDI or have lower growth rates, as the density gradient is more gradual.

For each set of parameters, we ran two types of simulations: an electrostatic (ES) case and an electromagnetic (EM) case. 

In the ES runs, we allowed only electrostatic fields (solving Poisson’s equation for \( \rho \)) and suppressed any perturbed magnetic field by not advancing Ampère’s law—effectively keeping \( \mathbf{B} \) fixed to its equilibrium value. These runs isolate the classic electrostatic LHDI behavior. While no magnetic reconnection in the strict topological sense can occur in this setup, we loosely refer to reconnection-like behavior when the current sheet evolves or bifurcates in response to electrostatic turbulence.

In the EM runs, we included the full coupling to Ampère’s law so that magnetic perturbations ($\delta B$) could grow self-consistently from the plasma currents; this captures the electromagnetic branch and any subsequent instability of the current sheet. 

All simulations were initialized with a small random perturbation (noise) in the electron distribution to seed the instability. We chose a noise amplitude low enough to remain in the linear regime initially, but sufficient to trigger the fastest growing LHDI modes. The typical grid resolution used was $N_y \times N_z = 128\times 64$ (for larger domains, up to $256 \times 64$ in some cases), and the velocity space for ions was discretized $N_{v_x}\times N_{v_y}\times N_{v_z}\times =32\times32\times32$ with and electrons was discretized with $N_{v_{\parallel}} =36$ (parallel velocity ) such that velocity-space resolution issues (e.g. trapping, Landau damping) were negligible. Time steps were chosen to resolve the lower hybrid period and the electron transit times (CFL condition for the Vlasov solver in phase space). Each simulation was run for several hundred $\Omega_{ci}^{-1}$, long enough to cover the linear growth and nonlinear saturation of LHDI, and (in EM cases) to observe any onset of reconnection or other secondary instabilities.

\section{Results}~\label{sec:results}
\begin{figure}[h]
    \centering
    \includegraphics[width=0.5\textwidth]{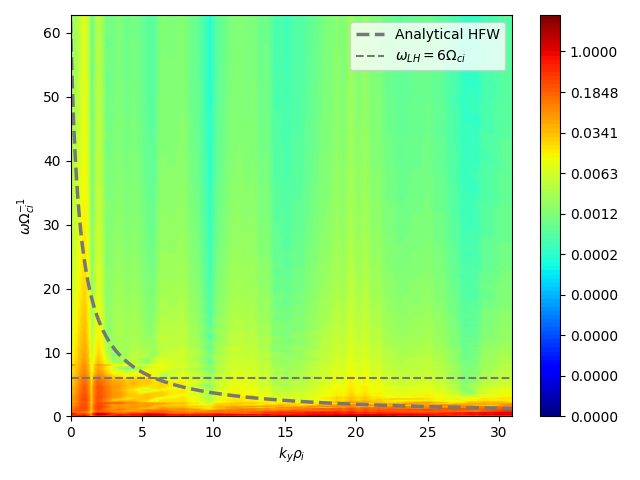} 
    \caption{Frequency-wave number spectrum of
electrostatic field (Ey) for mass ratio 36
(color coded), LHDW can be observed at $\omega_{LH} = 6 \Omega_{ci}$ and analytical dispersion
relation for the High Frequency Waves
(dashed line).
}
    \label{fig:lhw_hfw_spectrum}
\end{figure}
Figure~\ref{fig:lhw_hfw_spectrum} presents a contour plot of the perpendicular wavenumber ($k_y \rho_i$) versus the normalized frequency ($\omega/\Omega_{ci}$) spectrum obtained from simulations for mass ratio 36, Temperature ratio 10 and half thickness of the current sheet L= 1 $\rho_i$. The color scale represents the wave power intensity, with warmer colors indicating regions of enhanced fluctuation amplitude. Two key physical phenomena are successfully captured: lower hybrid drift waves (LHDWs) and high-frequency waves (HFWs) likely corresponding to Langmuir or upper-hybrid oscillations depending on local plasma parameters.

The lower-hybrid frequency normalized to the ion cyclotron frequency is
\begin{equation}
\frac{\omega_{LH}}{\Omega_{ci}} 
= \frac{\sqrt{\Omega_{ci}\Omega_{ce}}}{\Omega_{ci}} 
= \sqrt{\frac{\Omega_{ce}}{\Omega_{ci}}} 
= \sqrt{\frac{m_i}{m_e}},
\end{equation}
where we have used the relation 
$\Omega_{ce}/\Omega_{ci} = m_i/m_e$. 

This expression holds under standard assumptions for the electrostatic lower-hybrid frequency: a cold plasma with perpendicular wave propagation (\( \mathbf{k} \perp \mathbf{B} \)), in the regime \( \Omega_{ci} \ll \omega \ll \Omega_{ce} \), and neglecting finite temperature and displacement current effects.

For a proton–electron plasma (\( m_i/m_e = 36 \)), this yields
\[
\frac{\omega_{LH}}{\Omega_{ci}} \approx 6
\]
 
Turbulent wave activity at kinetic scales is of particular interest because a variety of wave modes—including whistler waves, kinetic Alfvén waves, and lower hybrid fluctuations—are believed to play key roles in energy dissipation in space plasmas. In the solar wind, for example, such turbulence-driven processes are thought to contribute significantly to plasma heating~\cite{Cranmer2002,Oughton2011,Podesta2012,Howes2011,Chen2016}.

In kinetic-scale turbulence, processes such as magnetic reconnection, 
Landau damping, and transit-time damping have been identified as key 
dissipation mechanisms~\cite{Hollweg1978,Karimabadi2014}. The hybrid kinetic--gyrokinetic formulation used 
here is capable of recovering high-frequency waves in the drift-kinetic limit, 
an important step toward modeling kinetic Alfvén wave (KAW)-like physics 
in the frequency range $\Omega_{ci} \lesssim \omega \lesssim \Omega_{ce}$, 
as discussed in~\cite{Lopes2022}.

The analytical HFW dispersion relation~\cite{Lopes2022}:
\begin{equation}
\begin{aligned}
\left(\frac{\omega}{\Omega_{ci}}\right)^2 
&= \frac{1}{2}\Bigg(
1 +
\frac{\tilde{\omega}_{pe}^2 k_\parallel^2 
      + \tilde{\omega}_{pi}^2 k^2}
     {c_v k^2 
      + \left(\frac{\tilde{\omega}_{pe}}
                    {\tilde{\Omega}_{cs}} 
        k_\perp\right)^2}
\Bigg) \\[6pt]
&\quad + \frac{1}{2}
\Bigg[
\Bigg(
\tilde{\Omega}_{cs}^2 +
\frac{\tilde{\omega}_{pe}^2 k_\parallel^2 
      + \tilde{\omega}_{ps}^2 k^2}
     {c_v k^2 
      + \left(\frac{\tilde{\omega}_{pe}}
                    {\tilde{\Omega}_{cs}} 
        k_\perp\right)^2}
\Bigg)^2 \\[4pt]
&\qquad\quad
+ \frac{4\left(\tilde{\omega}_{pe}^2 
             + \tilde{\omega}_{ps}^2\right)
        \tilde{\Omega}_{cs}^2 k_\parallel^2}
       {c_v k^2 
        + \left(\frac{\tilde{\omega}_{pe}}
                      {\tilde{\Omega}_{cs}} 
          k_\perp\right)^2}
\Bigg]^{1/2}
\end{aligned}
\label{eq:HFW_dispersion_norm}
\end{equation}
The analytical HFW dispersion relation uses the normalized plasma frequencies \( \tilde{\omega}_{pe} = \omega_{pe}/\Omega_{ci} \), \( \tilde{\omega}_{pi} = \omega_{pi}/\Omega_{ci} \), and the normalized ion-acoustic speed \( \tilde{c}_s = c_s/\Omega_{ci} \), where \( c_s = \sqrt{(T_e + T_i)/m_i} \) is the sound speed for a two-species (electron–ion) plasma.

The lower-hybrid wave (LHW) activity is evident near the lower-hybrid 
frequency, as indicated in Figure ~\ref{fig:lhw_hfw_spectrum} by the overlaid 
horizontal dashed line labeled $\omega_{LH} = 6\,\Omega_{ci}$. The simulation 
shows enhanced spectral power concentrated along this frequency band, in 
good agreement with theoretical expectations for the lower-hybrid drift 
instability. Additionally, Figure ~\ref{fig:lhw_hfw_spectrum} shows the simulation 
spectrum in the $(k_y\rho_i,\,\omega/\Omega_{ci})$ plane, where the overlaid 
curved dashed line corresponds to the analytical HFW dispersion relation 
derived from Eq.~\eqref{eq:HFW_dispersion_norm}. 

The simulation spectrum 
exhibits a clear enhancement of power along this analytical curve, indicating 
the presence of high-frequency wave activity consistent with the hybrid 
dispersion relation. The HFWs occupy the expected kinetic range and remain 
well-separated from the lower-hybrid branch, confirming that the hybrid 
kinetic formulation accurately captures high-frequency physics beyond the 
traditional gyrokinetic limit. These results demonstrate that the ssV code 
reliably resolves both the lower-hybrid and high-frequency wave dynamics 
in the hybrid kinetic regime, validating its capability to model multi-scale 
wave phenomena in magnetized plasmas with high fidelity.
\begin{figure}[ht]
    \centering
    \includegraphics[width=0.4\textwidth]{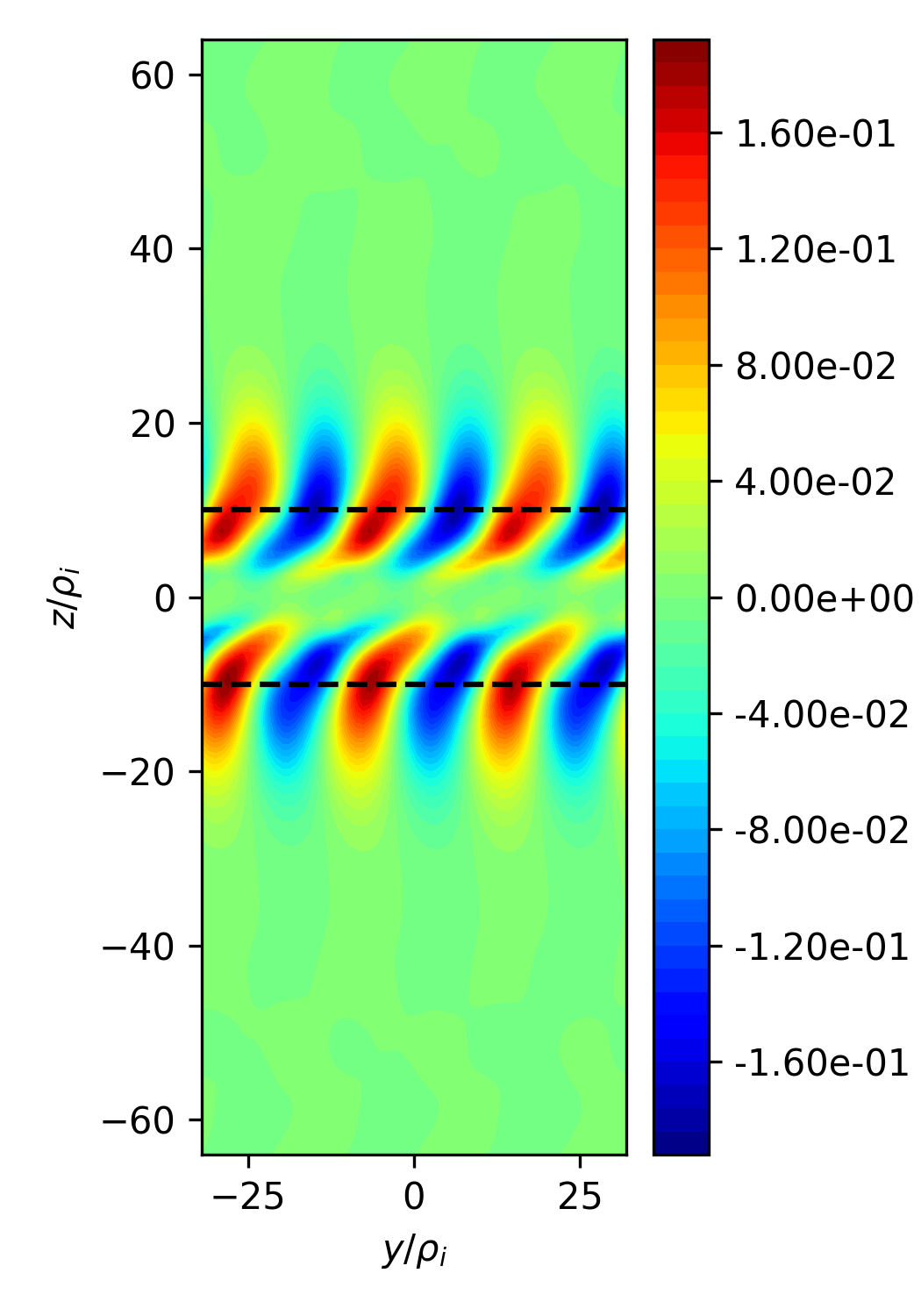}
    \caption{Electrostatic LHDI signatures in the electric field 
    $E_z$ for $m_i/m_e = 36$, $T_i/T_e = 10$, and 
    $L = 1\,\rho_i$ at $t \Omega_{ci}^{-1}=125$. Flute-like structures localized at the sheet (dotted black line) edges   
    ($z \approx \pm L$) are evident, consistent with the electrostatic 
    nature of the instability at low $\beta$.}
    \label{fig:ES_LHDI}
\end{figure}

\begin{figure}[ht]
    \centering
    \includegraphics[width=0.4\textwidth]{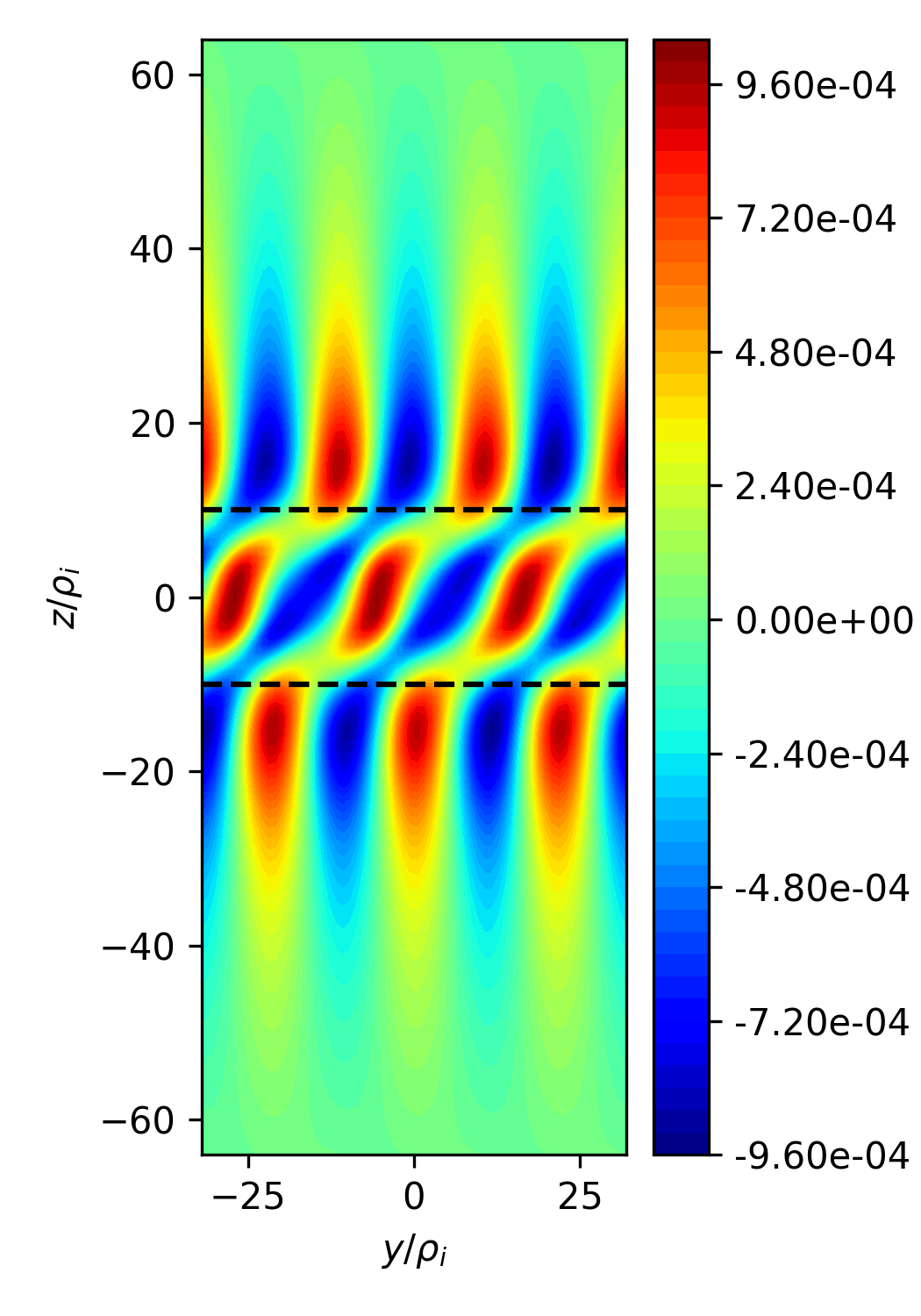}
    \caption{Electromagnetic LHDI signatures in the magnetic field $B_y$ 
    for $m_i/m_e = 36$, $T_i/T_e = 10$, $L = 1\,\rho_i$, and 
    plasma $\beta = 0.01$ at $t \Omega_{ci}^{-1}=125$. Long-wavelength undulations extending into the 
    sheet(dotted black line) center are visible, characteristic of the electromagnetic branch 
    of the LHDI.}
    \label{fig:EM_LHDI}
\end{figure}
Figures~\ref{fig:ES_LHDI} and~\ref{fig:EM_LHDI} show the characteristic 
signatures of the lower-hybrid drift instability (LHDI) in electrostatic (ES) 
and electromagnetic (EM) runs, respectively, for a simulation with 
$m_i/m_e = 36$, $T_i/T_e = 10$, and current sheet half-thickness 
$L = 1\,\rho_i$. In the ES case (Fig.~\ref{fig:ES_LHDI}), the LHDI is 
evident from the flute-like structure in the electrostatic electric field 
$E_z$, localized near the sheet edges ($z \approx \pm L$), consistent with 
the predominantly electrostatic nature of the instability at low $\beta$ 
\cite{daughton2003,Ricci2005}. In the EM case (Fig.~\ref{fig:EM_LHDI}), 
with plasma $\beta = 0.01$, the LHDI penetrates deeper into the current sheet, 
and the electromagnetic branch becomes active, as seen in the undulating 
patterns in the magnetic field $B_y$. These long-wavelength modulations are 
similar to the kinking structures reported in Vlasov and 10-moment simulations 
(Figs.~5 and~6 of ~\cite{innocenti2016,rahman2021}), indicating the transition to an 
electromagnetic regime at finite $\beta$.

\noindent\textbf{Linear Growth of LHDI:}
In all simulations, the lower hybrid drift instability was observed to initiate in the early phase as exponentially growing oscillations localized near the current sheet density gradient. 
\begin{figure}[h]
    \centering
    \includegraphics[width=0.5\textwidth]{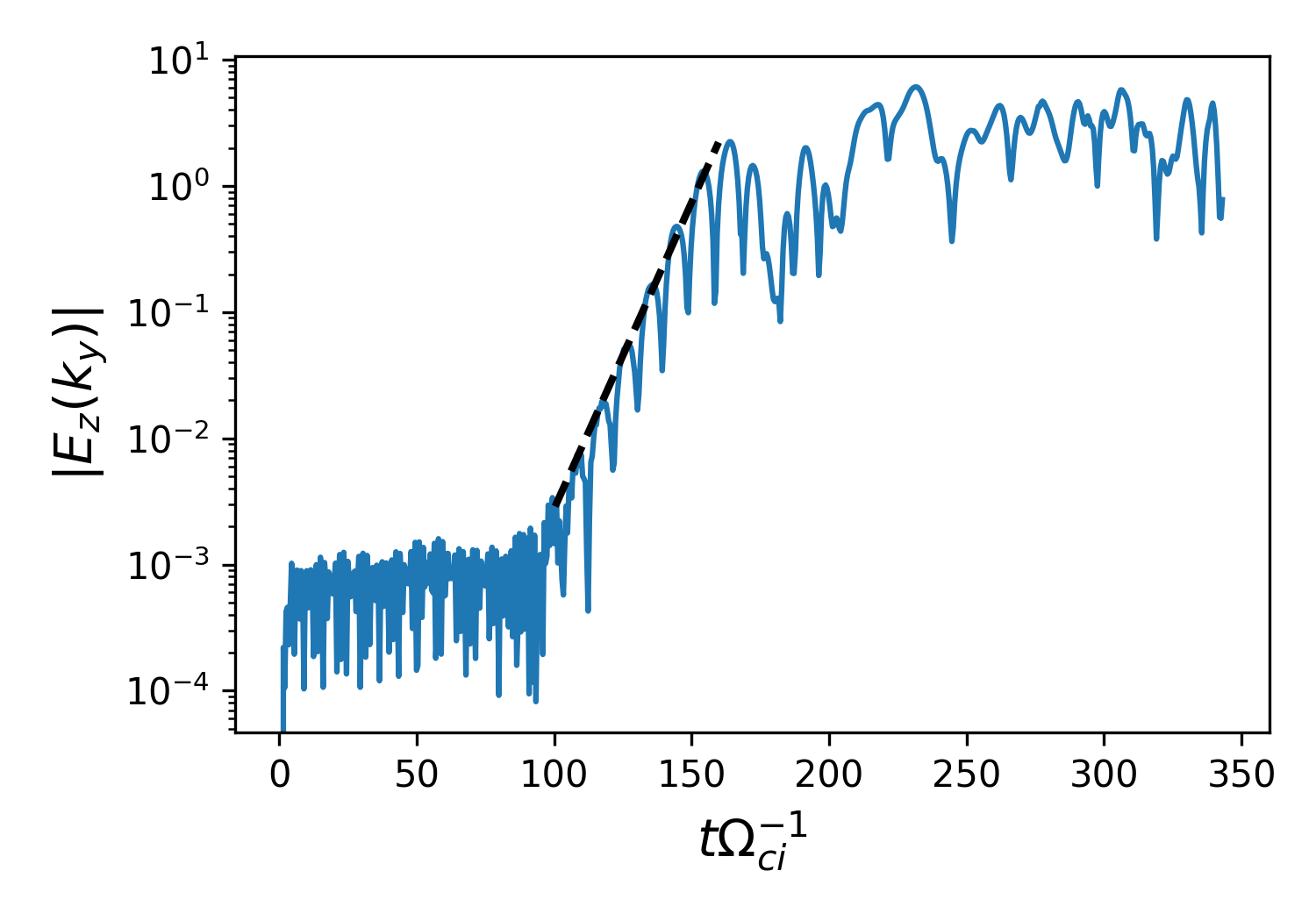} 
    \caption{Temporal evolution of the LHDI in a simulation run with 
ion-to-electron mass ratio $m_i/m_e = 36$, 
temperature ratio $T_i/T_e = 10$, 
and current sheet half-thickness $L = 1\,\rho_i$, 
as diagnosed by the amplitude of the electrostatic potential $\phi$
}
    \label{fig:temporaleval}
\end{figure}
Figure ~\ref{fig:temporaleval} shows a representative example of the Temporal evolution of the LHDI in a simulation run with 
ion-to-electron mass ratio $m_i/m_e = 36$, 
temperature ratio $T_i/T_e = 10$, 
and current sheet half-thickness $L = 1\,\rho_i$, 
as diagnosed by the amplitude of the electrostatic potential $\phi$ (or equivalently density perturbation) at $z \approx \pm 1\,\rho_i$ (sheet edge). A clear exponential growth is seen, from which a growth rate $\gamma$ can be extracted. We first examine how this linear growth rate depends on the key parameters. 

\begin{figure}[h]
    \centering
    \includegraphics[width=0.5\textwidth]{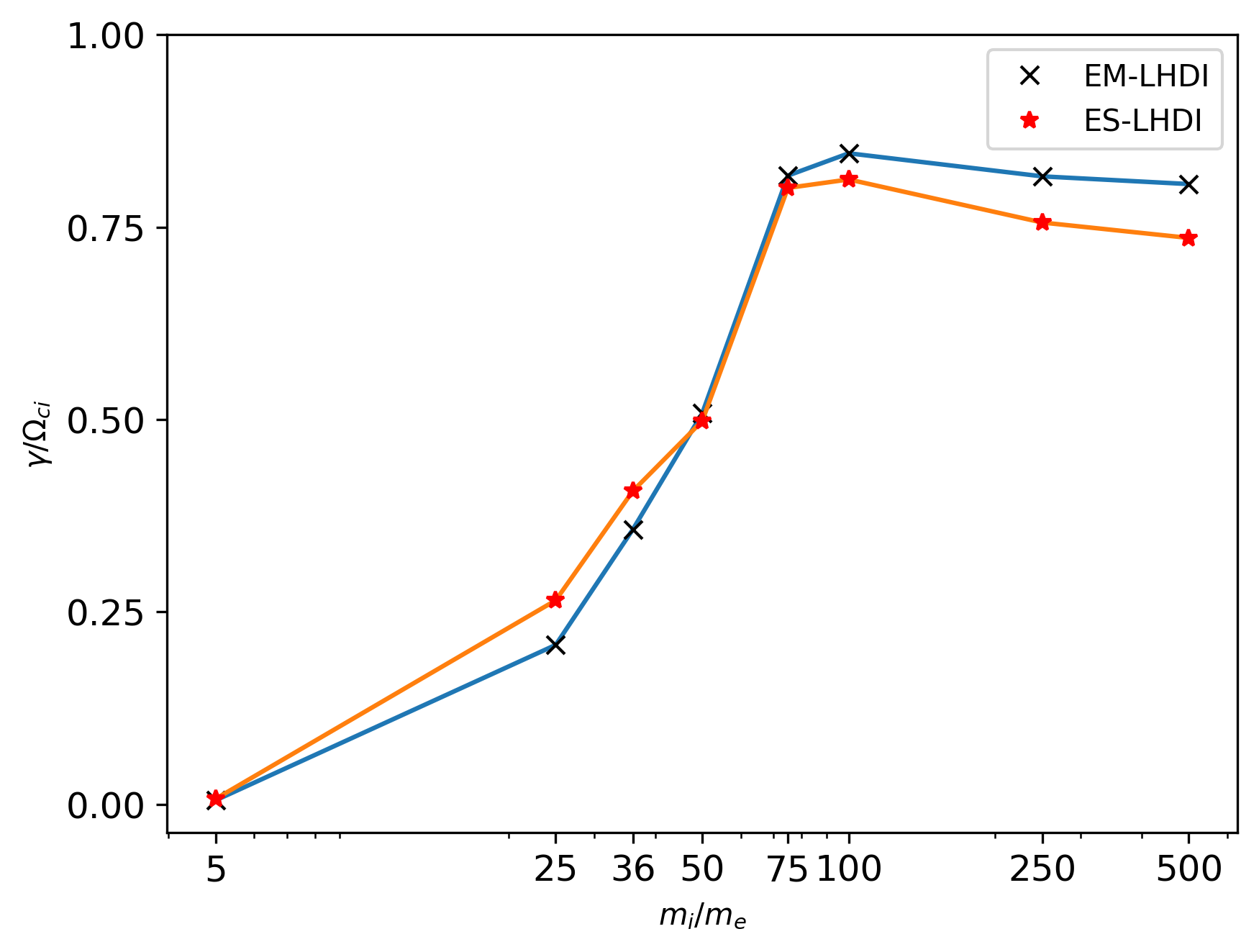} 
    \caption{Measured linear growth rates $\gamma/\Omega_{ci}$ as a function of $m_i/m_e$
}
    \label{fig:massratiocompare}
\end{figure}

\noindent\textbf{Mass ratio effect:} 
Figure~\ref{fig:massratiocompare} plots the measured linear growth rates $\gamma/\Omega_{ci}$ as a function of $m_i/m_e$, showing a steep rise that asymptotically approaches a plateau at high mass ratio. 
The growth rate of LHDI increases markedly with the ion-to-electron mass ratio $m_i/m_e$. For low mass ratios (e.g. 5 or 10), we found that the instability grows slowly and sometimes its amplitude stayed relatively low (nearly noise level for a long duration before eventually increasing). As $m_i/m_e$ is increased, the onset of rapid growth occurs earlier and $\gamma$ becomes larger. Quantitatively, going from $m_i/m_e=10$ to $100$ roughly doubled the measured growth rate in our simulations. By $m_i/m_e=250$, the growth rate begins to saturate in the sense that the increase from 100 to 500 is smaller than from 10 to 100. This trend is expected because a realistic mass ratio provides a clearer scale separation between electron and ion dynamics, allowing the classic LHDI physics to emerge more strongly. With too low mass ratio, electrons behave artificially “heavy” and do not drift as quickly relative to ions, weakening the drive of the instability. Our results are consistent with the notion that using a nearly physical mass ratio is necessary to accurately capture LHDI growth. In fact, a threshold behavior was observed: for $m_i/m_e < 10$ and certain parameter combinations, the LHDI was barely unstable or grew extremely slowly, whereas for $m_i/m_e \ge 50$ an abrupt jump in growth and saturation level occurred. This suggests that past simulations with very reduced mass ratios may have underestimated the role of LHDI. 

\begin{figure}[h]
    \centering
    \includegraphics[width=0.5\textwidth]{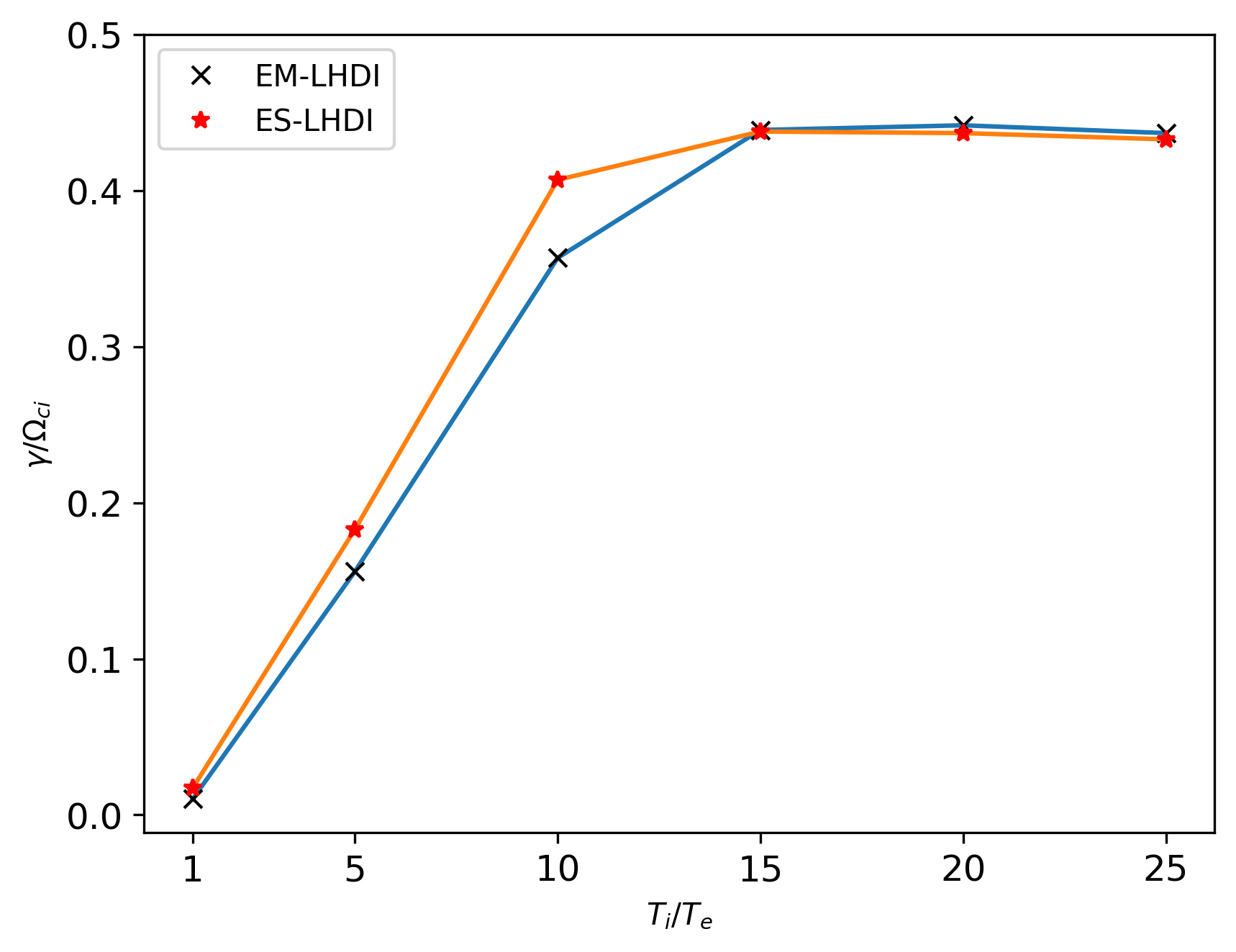} 
    \caption{Measured linear growth rates $\gamma/\Omega_{ci}$ as a function of $T_i/T_e$
}
    \label{fig:tempratiocompare}
\end{figure}

\noindent\textbf{Temperature ratio effect:} 
Figure~\ref{fig:tempratiocompare} shows the growth rate as a function of $T_i/T_e$ for a representative set of runs, illustrating the monotonic increase of $\gamma$ with temperature ratio.
The ion/electron temperature ratio $T_i/T_e$ was found to significantly influence the LHDI’s vigor. Generally, runs with higher $T_i/T_e$ exhibited faster growth and larger fluctuating fields. For example, at \( m_i/m_e = 36 \) and \( \omega_e = 0.01 \), increasing the temperature ratio from 1 (equal temperatures) to 25 led to a noticeably enhanced growth rate of the instability.
 The case $T_i/T_e = 1$ (ions and electrons equal temperature) had the weakest instability, sometimes requiring additional perturbation to get it to grow at all. This trend aligns with theoretical expectations: a larger $T_i/T_e$ means the ion pressure gradient (and hence ion diamagnetic drift) is stronger relative to the electron pressure gradient. Electrons, being cooler, are more tightly magnetized and have slower thermal motion, which means they can more readily E×B drift with the wave and be perturbed, enhancing the two-stream coupling that drives LHDI. In contrast, when the electron temperature becomes comparable to or exceeds the ion temperature (\( T_e \gtrsim T_i \)), the growth of the instability is reduced. This trend is primarily attributed to enhanced finite Larmor radius (FLR) damping and modifications to the diamagnetic drift contributions. As \( T_i/T_e \) decreases, the dominant \( k_\perp \) of the unstable mode tends to shift, further influencing the instability spectrum.
 Our simulations confirm that high ion temperature (or equivalently low electron temperature) is a favorable condition for LHDI. A side effect observed is that runs with large $T_i/T_e$ also showed more pronounced ion heating during LHDI saturation (since more free energy resides in the ion distribution; we discuss this later). 

\noindent\textbf{Current sheet thickness effect:} Perhaps the most sensitive parameter for LHDI is the current sheet thickness $L$. We find that thinner current sheets are dramatically more unstable to LHDI. In the thinnest case we tested ($L = 0.5 \rho_i$), the LHDI grew explosively fast – within a few ion gyroperiods – and reached large amplitude. For a moderate thickness $L = \rho_i$, growth was still significant, but for the thick case $L = 1.5 \rho_i$, the growth rate dropped substantially (by roughly an order of magnitude compared to $0.5 \rho_i$). In some $L=1.5 \rho_i$ runs, the LHDI initially grew so slowly that other perturbations (numerical noise or very low-growth modes) were competitive, leading to a more turbulent but low-level state rather than a clear LHDI domination. The physical reason is that a thicker sheet has a broader density gradient, so the driving diamagnetic current (proportional to $\nabla n$) is weaker. Moreover, for a given mode \( k_y \), a thicker current sheet places the mode further from resonance at the edges and may lead to increased damping in the central region. The trend we observe—higher growth rates for thinner sheets—is consistent with both linear theory and prior simulations~\cite{daughton2003, daughton2004, Ricci2005}.
 In fact, there appears to be a threshold sheet thickness (on the order of the ion gyroradius) below which the electromagnetic branch (mode B) of LHDI can arise and lead to dramatic effects on the current sheet, whereas above that thickness the instability remains mostly electrostatic and benign. In our EM runs, we indeed saw that only the thinnest sheets developed strong magnetic perturbations. The electrostatic runs still had LHDI for thin sheets, but for thick sheets (like $1.5 \rho_i$) the ES LHDI was very weak or nonexistent. This suggests that in practical terms, LHDI is likely to matter most in current sheets that have already thinned to near ion-scale – such as those during the late nonlinear phase of reconnection or in thin current layers in the magnetotail. 

\noindent\textbf{Plasma beta effect:} The plasma beta $\beta_e$ proved to affect primarily the nature of the LHDI fluctuations (electrostatic vs electromagnetic) rather than the existence of the instability itself. At very low beta ($\beta_e \sim 0.01$), the plasma is strongly magnetized; in these runs, the LHDI manifested almost entirely as electrostatic oscillations with negligible magnetic perturbation. The growth rate in the low-\( \beta \) runs remained moderate, consistent with previous studies showing that electrostatic LHDI can still develop in strongly magnetized plasmas~\cite{daughton2003,Ricci2005}. However, the saturation amplitude of \( \delta B \) was extremely small—indicating that the instability saturated by flattening the density profile, without significantly perturbing the magnetic field.
 As beta was increased, we observed that the electromagnetic contribution of LHDI grew. For $\beta_e \gtrsim 0.1$, especially in combination with thin sheets, the instability generated noticeable $\delta B_y$ and $\delta B_z$ perturbations in the EM simulations. These magnetic fluctuations were still small compared to the background field (at most a few percent of $B_0$ in our $\beta_e=0.2$ cases), but they were sufficient to deform the current sheet. Interestingly, the linear growth rates did not vary drastically across this 
$\beta$ range. A slight decrease in \( \gamma \) at the highest \( \beta \) is consistent with theoretical expectations and may result from stabilization mechanisms active at high \( \beta \)~\cite{daughton2003, Ricci2005}. This reduction could be attributed to a combination of factors, such as a decrease in the lower-hybrid frequency (e.g., due to higher density) or enhanced damping associated with warmer electrons or reduced magnetic field strength—each of which can contribute to increased \( \beta \).
 Physically, when $\beta_e$ is high, even a tiny current perturbation can create a noticeable magnetic perturbation (since the absolute magnetic field is lower). Conversely, at low beta the magnetic field is so strong that the same current fluctuation yields a minuscule fractional change in $B$. Thus, higher beta tends to favor the electromagnetic branch of LHDI. In summary, within $\beta_e=0.01$–0.2 we saw LHDI occur in all cases, but only the higher beta cases showed a coupled electromagnetic response. The electrostatic growth rate was relatively insensitive to beta, so the primary role of beta is enabling or disabling the $\delta B$ effects.

\noindent\textbf{Nonlinear Saturation and Current Sheet Modification:}
After several tens of inverse ion cyclotron times, the LHDI in each run enters a saturation phase where the exponential growth levels off. This saturation is achieved through different mechanisms in the ES and EM cases, with distinct consequences for the current sheet structure. In the electrostatic-only simulations, the LHDI saturates by flattening the plasma density and pressure profiles at the sheet edges. As the instability grows, it transports plasma across the magnetic field via E×B drift motions, effectively filling in the density trough at the sheet boundaries. 
\begin{figure}[h]
    \centering
    \includegraphics[width=0.5\textwidth]{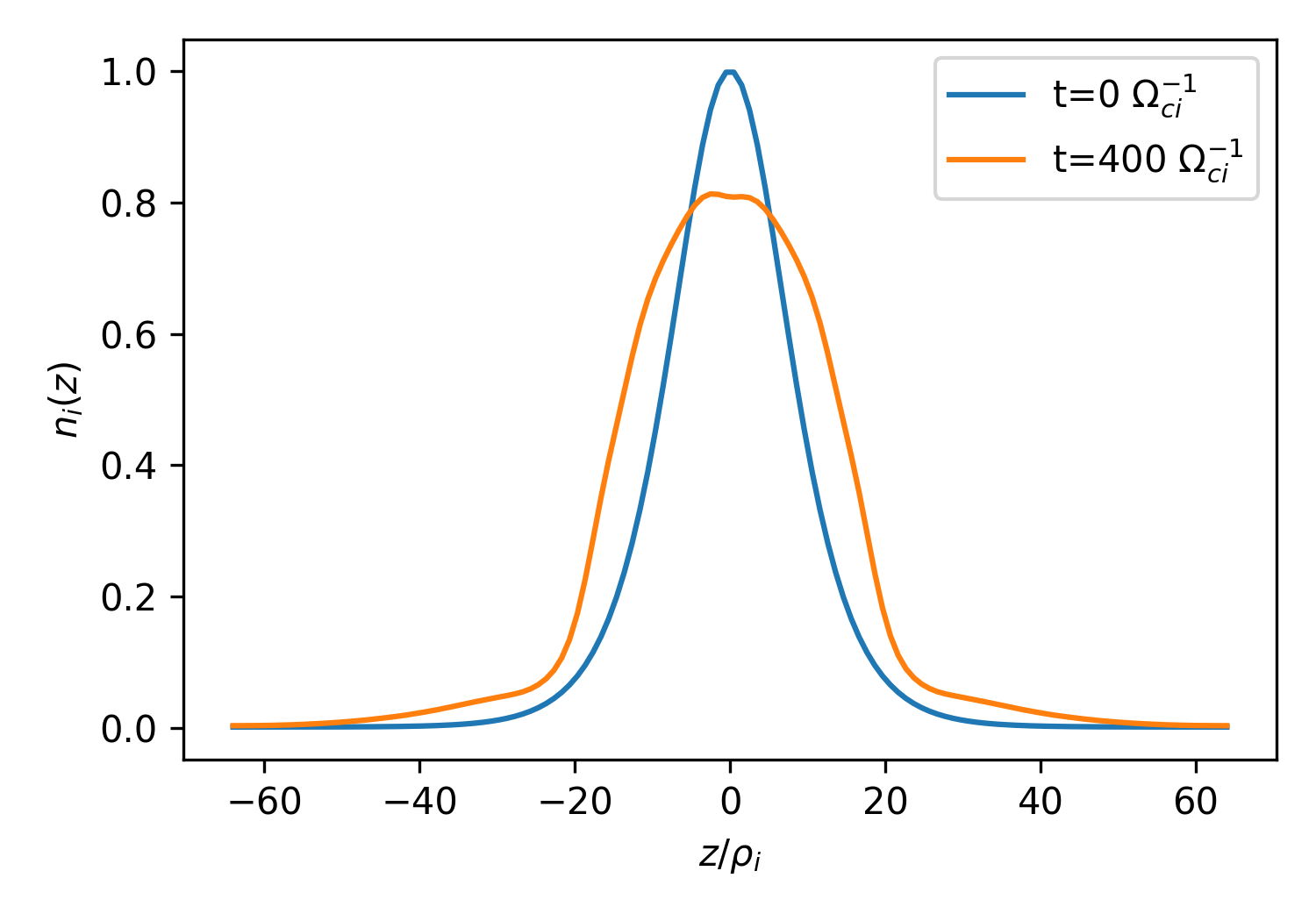} 
    \caption{Comparison of the initial and final density profile $n(z)$ for an electrostatic run with $L = 1 \rho_i$
}
    \label{fig:densitycompareES}
\end{figure}
Figure ~\ref{fig:densitycompareES} compares the initial and final density profile $n(z)$ for an electrostatic run with $L =1 \rho_i$. Initially, $n(z)$ peaks at the center and drops off toward the edges; after saturation, the edge density (at $|z| \sim L$) is noticeably higher (the gradient is reduced). This gradient reduction quenches the LHDI drive, since the free energy (pressure gradient) is largely exhausted. Associated with this, the electric potential fluctuations $\phi$ form a radial pattern of vortices that mix the plasma at the sheet periphery. The anomalous diffusion of particles flattens not only density but also the current density profile $J_x(z)$. We observe a slight broadening of the electron current layer in ES cases: the initial current, concentrated around $z=0$ with width $L$, spreads out a bit as electrons are scattered by the turbulent electrostatic fields. However, because no magnetic perturbations are allowed, the magnetic field profile \( B_y(z) \) remains fixed at its equilibrium configuration (i.e., it is not evolved during the simulation). There is no reconnection or topological change of field lines in purely electrostatic simulations; the \( B_y \) component remains antiparallel across the sheet, and a single current layer is maintained.
 Thus, in ES runs the LHDI’s effect is limited to creating a turbulent edge that increases particle mixing and slows the local drift velocities (a form of anomalous viscosity), but it does not directly cause the sheet to tear or reconnect. In the electromagnetic simulations, the nonlinear outcome is more dramatic. Once the LHDI grows to large amplitude, the small but finite magnetic perturbations ($\delta \mathbf{B}$) start to feed back on the current sheet equilibrium. We find that the saturated state in EM runs is characterized by a bifurcated and thinned current sheet. 
\begin{figure}[h]
    \centering
    \includegraphics[width=0.5\textwidth]{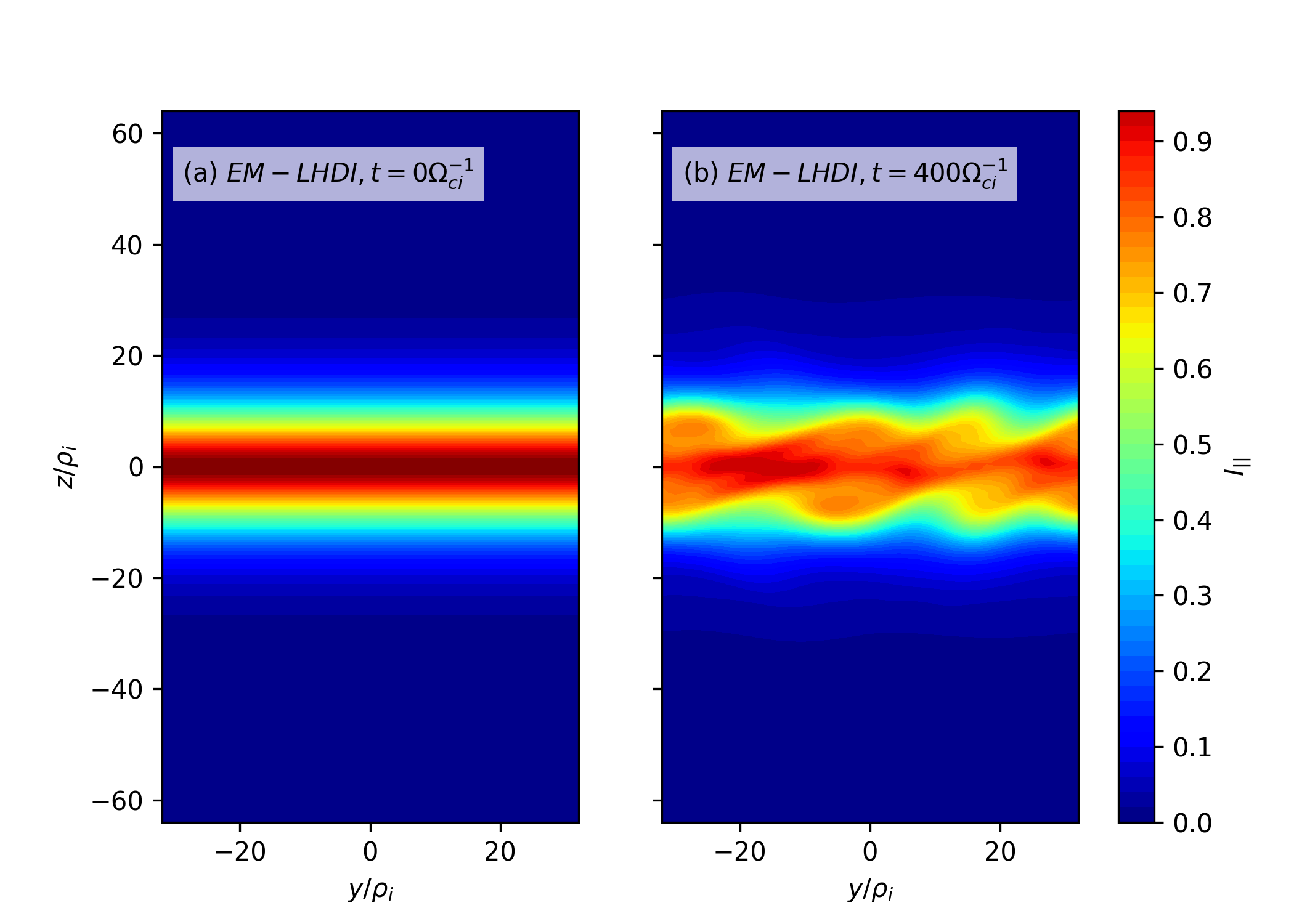} 
    \caption{Out-of-plane current density $J_x(z)$ before and after LHDI saturation in a representative EM run (with $L = \rho_i$, $m_i/m_e=36$, $\beta_e=0.01$)
}
    \label{fig:densitycompareEM}
\end{figure}
Figure~\ref{fig:densitycompareEM} shows the out-of-plane current density $J_x(z)$ before and after LHDI saturation in a representative EM run (with $L = \rho_i$, $m_i/m_e=36$, $\beta_e=0.01$). Initially, $J_x(z)$ has a single peak at $z=0$. After saturation, this single peak has split into two narrower peaks located slightly off-center (around $z \approx \pm 0.3,\rho_i$), with a noticeable drop in $J_x$ at $z=0$. In other words, the current sheet has bifurcated into a double-layer structure. Correspondingly, the magnetic field $B_y(z)$ develops a small plateau or bulge at the center (indicating a thinner effective current sheet). This behavior – current bifurcation – is a hallmark of nonlinear LHDI in kinetic simulations and was indeed observed in early implicit PIC studies of LHDI. The physical mechanism behind it in our simulations appears to be the following: As the LHDI flattens the density at the current sheet edges, the local 
pressure decreases there. This generates a pressure imbalance, with higher pressure at the sheet center exerting a net outward force toward the edges. If the edge pressure is sufficiently reduced and boundary constraints limit outward expansion, this imbalance can effectively squeeze the current layer inward, leading to localized compression and an increase in the central magnetic field gradient. This steepens \( \partial B / \partial z \), and by Ampère’s law, enhances \( J_x \) near the midplane. Alternatively, in more open configurations, the same pressure imbalance could drive plasma outward, resulting in a widened current sheet instead. The outcome likely depends on the global equilibrium and how the pressure profile evolves across the sheet.
 However, at the same time, the LHDI perturbations drive a significant electron cross-field motion that redistributes the current. Electrons get displaced from $z=0$ towards the sides (forming two distinct current channels), leaving a dent in the middle. The outcome is a thinner but double-peaked current profile. Notably, the total current integrated across the sheet remains roughly the same (since the imposed $B_0$ asymptotes far out haven’t changed much), but it is redistributed in space. 

Another striking effect observed in the electromagnetic (EM) cases is 
the onset of current sheet kinking—an undulating deformation of the sheet 
driven by electromagnetic modes that become prominent when the LHDI 
extends deeper into the current layer~\cite{innocenti2016,rahman2021}. This kinking manifests as 
long-wavelength distortions along the current direction and is often 
associated with the nonlinear evolution of the electromagnetic branch of 
the LHDI, particularly in high-$\beta$ or thin-sheet regimes where magnetic 
tension cannot fully suppress the instability~\cite{Ricci2005,daughton2004}. 
Such kinking has been linked to current sheet thinning, bifurcation, and 
even the onset of fast magnetic reconnection~\cite{daughton2004,Ricci2005,rahman2021}.

\noindent\textbf{Dependence on Parameters:}
Examining the results across our parameter scan, we can summarize the dependence of the LHDI nonlinear outcomes on mass ratio, temperature ratio, sheet thickness, and beta:

\noindent\textbf{Mass ratio:} Higher $m_i/m_e$ leads to more vigorous LHDI and a greater tendency for electromagnetic effects. At $m_i/m_e=5$ or 10, even in EM runs the current sheet showed only mild perturbation (the instability was so weak that it saturated by tiny density adjustments). At $m_i/m_e \ge 50$, the full spectrum of effects (density flattening, current bifurcation, kinking) was observed. This confirms that an adequate mass ratio is critical to simulate LHDI-driven phenomena~\cite{lin2008,rahman2021}. In intermediate cases (e.g. $m_i/m_e=25$), we saw LHDI grow but at a slower rate, and the magnetic perturbations took longer to appear; consequently, within the same physical time, a run with a low mass ratio might not yet develop a kink whereas a high mass ratio run would. This suggests caution when interpreting reduced-mass simulations of reconnection: they might under-predict the timing and impact of LHDI-related turbulence.

\noindent\textbf{Temperature ratio:} High $T_i/T_e$ not only increased growth rate but also resulted in stronger final turbulence levels. For example, with $T_i/T_e=20$ we observed more intense fluctuations and a more pronounced current sheet deformation than with $T_i/T_e=1$ (keeping other parameters same). In low $T_i/T_e$ runs, the electrons (being hotter) tended to smooth out some potential perturbations, and the LHDI saturated at a lower amplitude. Interestingly, we found that the threshold sheet thickness for electromagnetic mode development was somewhat higher when $T_i/T_e$ was small – i.e. with cold ions (and hot electrons), even a thin sheet did not produce as large a kink. This might be because hot electrons can carry current more effectively and stabilize the kink. In contrast, when ions are much hotter, the electrons (colder) are more easily frozen to field lines, and the current layer is more susceptible once perturbed. Thus, large $T_i/T_e$ exacerbates the LHDI’s impact on the current sheet.

\noindent\textbf{Sheet thickness:} As described, $L$ had a qualitative effect. For $L \lesssim \rho_i$, we invariably saw substantial LHDI effects including electromagnetic perturbations and sometimes secondary instabilities. For $L > \rho_i$, the instability was marginal; in some $L=2 \rho_i$ runs with high $m_i/m_e$, we saw a brief initial LHDI growth that then saturated at very low amplitude (essentially flattening a very slight gradient) and nothing further happened – the sheet remained intact and little turbulence persisted. There appears to be a critical thickness around $L \sim 1$–$1.5 \rho_i$ dividing these regimes, which aligns with theoretical expectations for mode B onset. Near this threshold, the combination of LHDI and tearing becomes important: if a sheet is just slightly thicker than the threshold, LHDI alone might not break it, but if a tearing mode or external perturbation begins to thin it, LHDI could then kick in strongly. Our data suggest a synergistic scenario where thinning and LHDI amplify each other once $L$ falls below $\sim \rho_i$.

\noindent\textbf{Beta:}  Low-beta cases remained primarily electrostatic, while higher-beta cases exhibited significant electromagnetic activity. In the highest beta run ($\beta_e=0.2$) with a thin current sheet, the LHDI produced strong electric field fluctuations that, based on prior kinetic studies~\cite{daughton2003, huba1977}, are expected to generate nonlinear electron phase-space structures (e.g., trapped particle regions or electron holes). These structures are characteristic of electrostatic wave saturation and can lead to flattening of the electron distribution function, thereby limiting further growth of the instability.  In lower beta runs, damping by electrons was less significant (since $T_e$ was lower for lower beta given fixed pressure) and we saw more fluid-like saturation (via density flattening). Overall, beta influenced how the energy split between fields and particles: higher beta gave more energy in magnetic fields (thus affecting the topology), whereas low beta put more into particle thermal energy.

A final important result is that the purely electrostatic vs fully electromagnetic simulations differ substantially in their end state. 
\begin{figure}[h]
    \centering
    \includegraphics[width=0.5\textwidth]{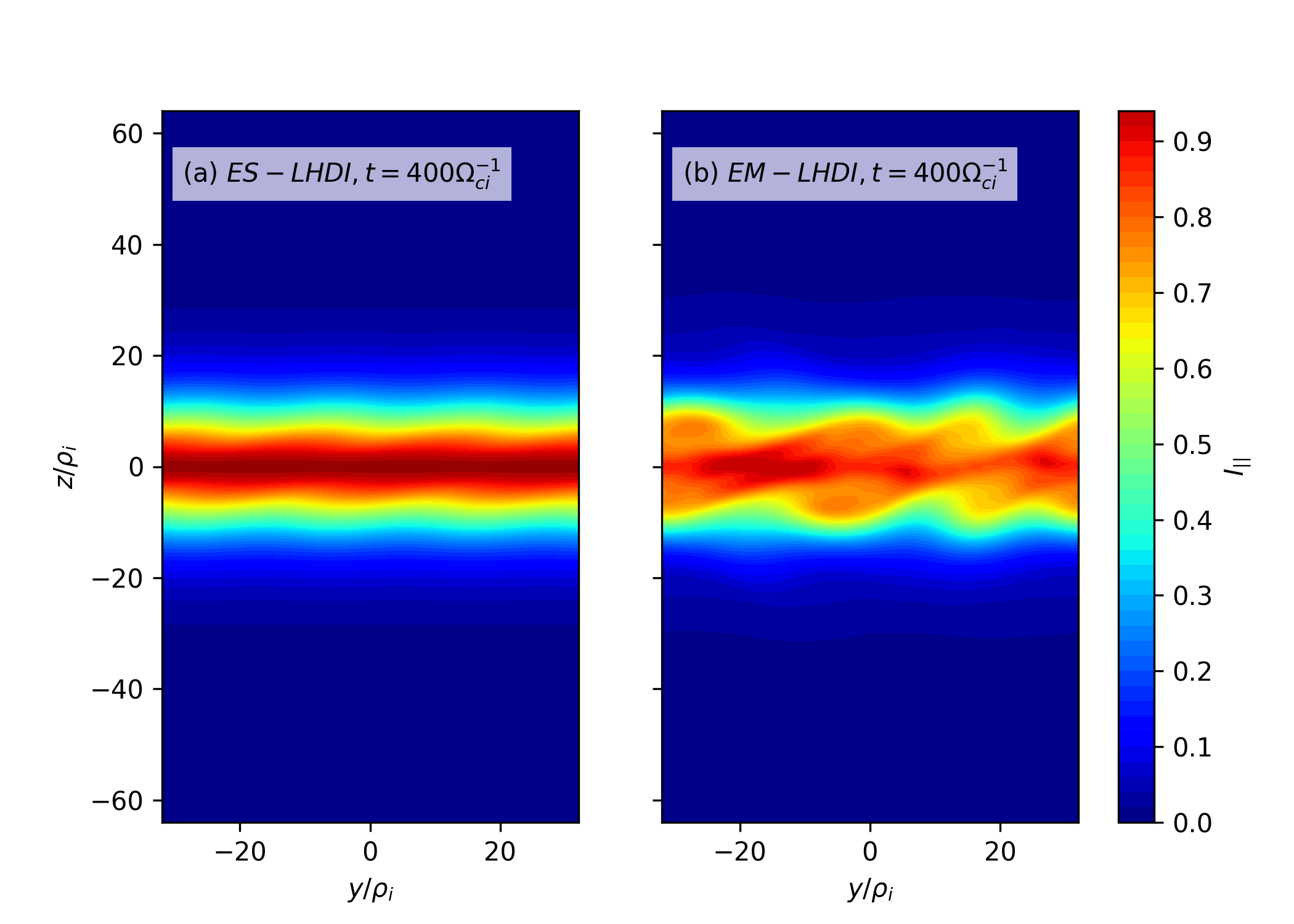} 
    \caption{Current sheet profile of ES and EM simulations runs with same initial conditions
}
    \label{fig:currentsheetcompareEM}
\end{figure}
Figure~\ref{fig:currentsheetcompareEM} directly contrasts an ES and an EM simulation that start from the same initial condition (for instance, $m_i/m_e=36$, $L=\rho_i$, $T_i/T_e=10$, $\beta_e=0.01$). In the ES case, after saturation, the current sheet is essentially still one layer (slightly broadened), and no reconnection or magnetic field change occurred. In the EM case, at the same time, the current sheet is bifurcated and a small magnetic island is visible at the center (indicating reconnection took place). This comparison underscores that while electrostatic LHDI can cause significant turbulence, only when electromagnetic effects are included can it drastically restructure the current sheet and influence reconnection. In a sense, the electrostatic LHDI is self-limiting – it saturates by smoothing the gradients that drive it – whereas the electromagnetic LHDI allows a positive feedback: as it thins the sheet, the central current intensifies which can further drive reconnection. Thus, to fully assess LHDI’s role in plasma dissipation and reconnection, electromagnetic effects must be considered.

\section{Discussion}\label{sec:discussion}
\noindent\textbf{Dual Role of LHDI in Current Sheet Evolution:}
Our numerical findings can be interpreted in the context of existing theoretical predictions and previous simulation studies of LHDI and reconnection. Broadly, we find a two-fold role of the LHDI in current sheet evolution: (1) it enhances plasma transport (anomalous resistivity and viscosity), tending to relax sharp gradients, and (2) it can destabilize the current sheet (via thinning and kinking), potentially accelerating the onset of reconnection. Which of these behaviors dominates appears to depend on how close the system is to the threshold of tearing instability. 

\noindent\textbf{Parameter Dependence: Sheet Thickness and Mass Ratio:}
In cases where the current sheet was relatively thick/stable, the LHDI in our simulations primarily acted as a benign turbulence that diffused the plasma. The instability saturated by smoothing out the density and pressure profile, consistent with the classic concept of LHDI-driven anomalous resistivity~\cite{daughton2003,Ricci2005}: electron scattering off the lower hybrid waves effectively produces a friction that opposes the current, thereby flattening the current profile. In these runs, we observed no significant change in the overall reconnection rate (indeed no reconnection was triggered at all). This aligns with the conclusions of some prior works that LHDI does not substantially alter the reconnection rate in and of itself. For example, Allmann-Rahn et al~\cite{rahman2021} found in a fluid-model study that including LHDI-induced effects did not significantly speed up reconnection, but it did cause enhanced anomalous transport (e.g. relaxed density gradients). 

Our kinetic results reinforce that viewpoint for the parameter regime of thicker sheets or lower mass ratios: the turbulence generated by LHDI redistributes plasma and current without initiating fast reconnection. In fluid terms, one might say LHDI provides an anomalous viscosity, as it facilitates momentum transfer across the field (e.g. the ion flow at the sheet edges is braked and momentum is transferred to background plasma). This viscosity tends to stabilize the sheet against other instabilities by erasing the free energy gradients. In the case of thin current sheets (on ion scales), the simulations exhibited current sheet thinning and bifurcation in the presence of strong LHDI activity. These structural changes are consistent with mechanisms proposed in earlier kinetic studies, where LHDI-induced modifications were suggested to facilitate or trigger fast reconnection by altering current sheet stability~\cite{daughton2004,Ricci2005}.

\noindent\textbf{Reconnection Triggering in Electromagnetic Regimes:}
In our electromagnetic runs for $L \lesssim \rho_i$, the LHDI-induced collapse of the current sheet produced conditions—such as intensified current density and enhanced turbulence—that are theoretically favorable for electron heating and for the rapid growth of the collisionless tearing mode. Although our simulation domain was not optimized for observing a sustained reconnection (we did not, for instance, include a long length in $x$ for a full X-line and exhaust to form), the emergence of a magnetic island in some runs is a smoking gun that LHDI drove the system past the reconnection threshold. We can compare this with Daughton et al~\cite{daughton2004} who reported that in their PIC simulations a “dramatic enhancement” of the central current by LHDI led to rapid reconnection onset. Our hybrid model reproduces the qualitative effect they described: the current sheet in our case bifurcated and thinned under LHDI, analogous to the “strong bifurcation of the current density” in their work. This supports the theoretical view that kinetic lower hybrid turbulence may act as a trigger for reconnection in nearly collisionless plasma sheets. While we do not quantify the reconnection rate in this study, our results suggest that LHDI contributes to conditions favorable for reconnection onset—specifically by thinning and bifurcating the current sheet more rapidly than collisional processes would. The simulations indicate that LHDI can accelerate the approach to the tearing threshold, even if the full development of a steady-state reconnection outflow lies beyond the simulated timescale. 

\noindent\textbf{Anomalous Resistivity vs. Anomalous Viscosity:}
In practical terms, for space plasmas like Earth’s magnetotail, this suggests that intense lower hybrid wave activity could precede and herald a substorm onset by making the current sheet more susceptible to tearing. The interplay between anomalous resistivity and anomalous viscosity in our results is also worth discussing. Anomalous resistivity refers to the effective resistivity due to turbulent scattering of current-carrying electrons, which enables violation of Ohm’s law $E = \eta J$. Anomalous viscosity refers to the turbulent momentum transport that can reduce flow shear (affecting ion flows and pressure balance). 

The LHDI primarily interacts with electrons (its phase velocity is typically of order the electron thermal or drift speed, so electrons resonate with it), thus one expects it to contribute mainly to anomalous resistivity. Theoretically, the electron velocity distribution may become isotropized and flattened due to wave-particle interactions, implying momentum exchange that would manifest as a non-ideal term in the generalized Ohm’s law. However, in the nonlinear stage, LHDI also produced large-scale perturbations (like the kink) that involve ion inertia and momentum – here the effect is like an anomalous viscosity, as the entire ion current layer is wiggling and redistributing momentum to the background fields. Our simulations show both effects: initially, electron scattering (resistivity) dominates saturation; subsequently, the sheet motion (viscous-like momentum transport, as in a kink) takes over. Distinguishing their contributions in a single simulation is challenging, but we can qualitatively say: in thick sheet cases, viscosity dominates (smooth shear, no reconnection), in thin sheet cases, resistivity becomes important (breaking field lines at X-line). 

\noindent\textbf{Mode Structure and Growth Sequence: Comparison to Theory:}
This is likely due to the turbulent Hall electric fields generated by LHDI scattering of electrons. Our results compare favorably with the theoretical framework put forth by Innocenti et al. (2016) for LHDI in thin sheets. They described modes A, B, C and noted that mode B (electromagnetic LHDI) develops in thin sheets and has growth rate on order $\Omega_{ci}$, and then kink-like modes follow. In our simulations, we indeed see a two-stage development: first a fast edge mode (A) saturates, then a slower electromagnetic perturbation (B) grows in the center, followed by a kink (C). The time sequence and spatial structure match their description qualitatively. Additionally, our finding that realistic mass ratio is needed echoes Innocenti’s~\cite{innocenti2016} point that high $m_i/m_e$ is “essential to ensure clear separation between electron and ion scales” – in lower mass ratio runs, the mode B was likely not well-resolved or might merge with mode A, altering the nonlinear outcome. This underscores that hybrid models like ours are valuable: they allow pushing toward higher mass ratios than fully kinetic codes typically can, making the results more directly relevant to physical plasmas. 

\noindent\textbf{Limitations of the Present Study:}
One should also consider the limitations of the present study. We used a 2D model (with a small third dimension conceptually for LHDI wavevector). True 3D physics might introduce additional couplings; for example, in a 3D system with both $k_x$ and $k_y$ variations, the turbulence could become more isotropic and perhaps less coherent in producing a single kink. Also, the drift-kink instability itself is inherently 3D (it requires an extended current in the out-of-plane direction to kink). Our simulation likely captured a mix of drift-kink and lower hybrid modes through the imposed periodicity in $x$, but a dedicated 3D run would allow $k_y$ to be a continuous spectrum and possibly show a cascade of modes rather than one dominant wavelength. The fluid 10-moment simulations of Allmann-Rahn et al~\cite{rahman2021}. did find that fully 3D reconnection with LHDI present leads to a turbulent state where many modes (LHDI, kink, etc.) coexist. They observed that the LHDI’s saturation level was higher than expected and it strongly kinked the sheet, consistent with what we see when allowing EM coupling. They also noted the dependence on initial perturbation amplitude for how turbulent it gets, which in our case is analogous to the fact that a sufficiently thin sheet (effectively a large perturbation from stable MHD) was needed to go turbulent. 

\noindent\textbf{Connections to Space and Laboratory Observations:}
We can also compare to laboratory experiments and space observations. The magnetopause observations often show intense lower hybrid waves at the edges of reconnection current layers, but these are usually believed not to significantly alter the reconnection rate – rather, they cause cross-field diffusion of magnetosheath plasma into the magnetospheric side. This scenario is akin to our moderate cases: LHDI at edges, mixing plasma but reconnection largely governed by other physics (like Hall fields). In contrast, the magnetotail thinner current sheets (during substorms) might be an environment where LHDI could actually push the sheet to reconnect. Some recent Magnetospheric Multiscale (MMS) observations have hinted at lower hybrid range fluctuations inside the diffusion region, which could be playing a role in electron scattering. Our simulation suggests that if the current sheet is already marginal, LHDI turbulence inside it (especially electromagnetic fluctuations) can indeed assist reconnection by scattering electrons and enabling the field lines to break more easily. For fusion plasmas (tokamak edge or sawtooth crashes), LHDI might contribute to anomalous transport along with other drift waves; its direct role in reconnection (e.g. sawtooth reconnection) is less studied, but in principle the same physics could apply if very steep gradients develop. 

\noindent\textbf{Synthesis and Implications for Reconnection Onset:}
In conclusion, our hybrid kinetic simulations provide a coherent picture that reconciles earlier seemingly contradictory views: LHDI alone will not “cause” reconnection in a stable sheet (it saturates by smoothing edges), but in a sheet that is near critical (thin current sheet), LHDI can dramatically accelerate the reconnection process by creating conditions favorable for the tearing mode to explosively develop. In the latter case, one could say anomalous resistivity (from LHDI) triggers reconnection; in the former case, reconnection likely needs to be initiated by some other means (e.g. an external perturbation or another instability like the electron tearing mode), and LHDI then merely accompanies it, contributing to turbulence and energy dissipation without changing the reconnection rate much. The relationship may also be synergistic: as reconnection begins and the sheet thins, LHDI will get stronger, which in turn could further enhance reconnection – a positive feedback loop, up to the point where gradients are flattened and a turbulent equilibrium is reached. Our results encourage future studies to explore this feedback in more detail (for instance, by measuring reconnection electric fields in simulations with and without LHDI, or by controlling the level of turbulence).

\section{Conclusion}\label{sec:conclusion}

We have investigated the nonlinear evolution of the lower hybrid drift instability (LHDI) in a collisionless reconnecting current sheet using a hybrid kinetic model with fully kinetic ions and drift-kinetic electrons. By systematically varying key plasma parameters—including mass ratio, temperature ratio, electron beta, and current sheet thickness—we explored the onset and effects of both electrostatic and electromagnetic LHDI.

\noindent\textbf{LHDI scaling and saturation behavior:}  
Our results confirm that the growth and structure of LHDI are highly sensitive to the ion-electron mass ratio and current sheet thickness. Realistic mass ratios and thin sheets (thickness $\lesssim \rho_i$) yield strong LHDI activity, while thicker sheets exhibit weaker, edge-localized modes. In electrostatic regimes, LHDI saturates via edge turbulence without altering the global current structure. In electromagnetic cases, the instability drives sheet bifurcation and kinking, consistent with kinetic theories describing electromagnetic LHDI (mode B) and subsequent kink modes (mode C).

\noindent\textbf{Reconnection triggering and anomalous transport:}  
We find that LHDI can precipitate reconnection by thinning and bifurcating the sheet, particularly in marginally stable configurations. In thicker sheets, it acts primarily as an anomalous viscosity, flattening profiles and enhancing cross-field transport. The instability thus plays a conditional role in reconnection onset, depending on background gradients and sheet thickness.

\noindent\textbf{Electron heating and wave-particle interactions:}  
LHDI turbulence leads to significant perpendicular electron heating and moderate ion heating. These heating signatures reflect the underlying wave-particle resonances and contribute to the energy budget of the current sheet.

\noindent\textbf{Implications for modeling and observations:}  
The ssV hybrid model effectively captured these dynamics at lower computational cost than full PIC simulations, supporting its use in multi-scale studies. Our findings reinforce the relevance of LHDI in both space and laboratory plasmas, particularly in explaining observed current sheet turbulence and heating in thin-sheet environments like Earth’s magnetotail. For global models, incorporating LHDI-driven anomalous resistivity or transport closures may be critical for accurately predicting reconnection onset.

\noindent\textbf{Outlook:}  
Future work should extend these simulations to 3D geometries and include additional kinetic effects, particularly for electrons. Exploring the interaction of LHDI with other instabilities (e.g., Kelvin-Helmholtz or electron-ion hybrid modes) would further illuminate its role in reconnection onset and turbulence.

In summary, LHDI emerges as a key driver of turbulence, heating, and possibly fast reconnection in collisionless current sheets. Capturing its effects accurately is essential for realistic modeling of space and laboratory plasma dynamics.

\begin{acknowledgments}
The authors gratefully acknowledge support from the Helmholtz Young Investigator
Group grant VH-NG-1239. Computational resources were provided by the
MPCDF Center, Garching. We also extend our gratitude to the Theoretical Physics
Department at Ruhr University Bochum for their collaboration and for providing the
base version of the MUPHY I code, which served as the foundation for ssV development.
\end{acknowledgments}

\section*{Data Availability Statement}
The data that support the findings of this study are available from the corresponding author upon reasonable request.

\bibliography{aipsamp}

\begin{thebibliography}{37}%
\makeatletter
\providecommand \@ifxundefined [1]{%
 \@ifx{#1\undefined}
}%
\providecommand \@ifnum [1]{%
 \ifnum #1\expandafter \@firstoftwo
 \else \expandafter \@secondoftwo
 \fi
}%
\providecommand \@ifx [1]{%
 \ifx #1\expandafter \@firstoftwo
 \else \expandafter \@secondoftwo
 \fi
}%
\providecommand \natexlab [1]{#1}%
\providecommand \enquote  [1]{``#1''}%
\providecommand \bibnamefont  [1]{#1}%
\providecommand \bibfnamefont [1]{#1}%
\providecommand \citenamefont [1]{#1}%
\providecommand \href@noop [0]{\@secondoftwo}%
\providecommand \href [0]{\begingroup \@sanitize@url \@href}%
\providecommand \@href[1]{\@@startlink{#1}\@@href}%
\providecommand \@@href[1]{\endgroup#1\@@endlink}%
\providecommand \@sanitize@url [0]{\catcode `\\12\catcode `\$12\catcode
  `\&12\catcode `\#12\catcode `\^12\catcode `\_12\catcode `\%12\relax}%
\providecommand \@@startlink[1]{}%
\providecommand \@@endlink[0]{}%
\providecommand \url  [0]{\begingroup\@sanitize@url \@url }%
\providecommand \@url [1]{\endgroup\@href {#1}{\urlprefix }}%
\providecommand \urlprefix  [0]{URL }%
\providecommand \Eprint [0]{\href }%
\providecommand \doibase [0]{http://dx.doi.org/}%
\providecommand \selectlanguage [0]{\@gobble}%
\providecommand \bibinfo  [0]{\@secondoftwo}%
\providecommand \bibfield  [0]{\@secondoftwo}%
\providecommand \translation [1]{[#1]}%
\providecommand \BibitemOpen [0]{}%
\providecommand \bibitemStop [0]{}%
\providecommand \bibitemNoStop [0]{.\EOS\space}%
\providecommand \EOS [0]{\spacefactor3000\relax}%
\providecommand \BibitemShut  [1]{\csname bibitem#1\endcsname}%
\let\auto@bib@innerbib\@empty
\bibitem [{\citenamefont {Daughton}(2003)}]{daughton2003}%
  \BibitemOpen
  \bibfield  {author} {\bibinfo {author} {\bibfnamefont {W.}~\bibnamefont
  {Daughton}},\ }\bibfield  {title} {\enquote {\bibinfo {title}
  {Electromagnetic properties of the lower-hybrid drift instability in a thin
  current sheet},}\ }\href {\doibase 10.1063/1.1594724} {\bibfield  {journal}
  {\bibinfo  {journal} {Physics of Plasmas}\ }\textbf {\bibinfo {volume}
  {10}},\ \bibinfo {pages} {3103--3119} (\bibinfo {year} {2003})}\BibitemShut
  {NoStop}%
\bibitem [{\citenamefont {Daughton}, \citenamefont {Scudder},\ and\
  \citenamefont {Karimabadi}(2006)}]{daughton2004}%
  \BibitemOpen
  \bibfield  {author} {\bibinfo {author} {\bibfnamefont {W.}~\bibnamefont
  {Daughton}}, \bibinfo {author} {\bibfnamefont {J.}~\bibnamefont {Scudder}}, \
  and\ \bibinfo {author} {\bibfnamefont {H.}~\bibnamefont {Karimabadi}},\
  }\bibfield  {title} {\enquote {\bibinfo {title} {Nonlinear evolution of the
  lower-hybrid drift instability in a current sheet},}\ }\href {\doibase
  10.1063/1.2218817} {\bibfield  {journal} {\bibinfo  {journal} {Physics of
  Plasmas}\ }\textbf {\bibinfo {volume} {13}},\ \bibinfo {pages} {072101}
  (\bibinfo {year} {2006})}\BibitemShut {NoStop}%
\bibitem [{\citenamefont {Hesse}\ and\ \citenamefont
  {Cassak}(2019)}]{hesse2019}%
  \BibitemOpen
  \bibfield  {author} {\bibinfo {author} {\bibfnamefont {M.}~\bibnamefont
  {Hesse}}\ and\ \bibinfo {author} {\bibfnamefont {P.~A.}\ \bibnamefont
  {Cassak}},\ }\bibfield  {title} {\enquote {\bibinfo {title} {Magnetic
  reconnection in the space sciences: Past, present, and future},}\ }\href
  {\doibase 10.1029/2018JA025935} {\bibfield  {journal} {\bibinfo  {journal}
  {Journal of Geophysical Research: Space Physics}\ }\textbf {\bibinfo {volume}
  {125}} (\bibinfo {year} {2019}),\ 10.1029/2018JA025935}\BibitemShut {NoStop}%
\bibitem [{\citenamefont {Yamada}, \citenamefont {Kulsrud},\ and\ \citenamefont
  {Ji}(2010)}]{yamada2010}%
  \BibitemOpen
  \bibfield  {author} {\bibinfo {author} {\bibfnamefont {M.}~\bibnamefont
  {Yamada}}, \bibinfo {author} {\bibfnamefont {R.}~\bibnamefont {Kulsrud}}, \
  and\ \bibinfo {author} {\bibfnamefont {H.}~\bibnamefont {Ji}},\ }\bibfield
  {title} {\enquote {\bibinfo {title} {Magnetic reconnection},}\ }\href
  {\doibase 10.1103/RevModPhys.82.603} {\bibfield  {journal} {\bibinfo
  {journal} {Reviews of Modern Physics}\ }\textbf {\bibinfo {volume} {82}},\
  \bibinfo {pages} {603--664} (\bibinfo {year} {2010})}\BibitemShut {NoStop}%
\bibitem [{\citenamefont {Zweibel}\ and\ \citenamefont
  {Yamada}(2009)}]{zweibel2009}%
  \BibitemOpen
  \bibfield  {author} {\bibinfo {author} {\bibfnamefont {E.~G.}\ \bibnamefont
  {Zweibel}}\ and\ \bibinfo {author} {\bibfnamefont {M.}~\bibnamefont
  {Yamada}},\ }\bibfield  {title} {\enquote {\bibinfo {title} {Magnetic
  reconnection in astrophysical and laboratory plasmas},}\ }\href {\doibase
  10.1146/annurev-astro-082708-101726} {\bibfield  {journal} {\bibinfo
  {journal} {Annual Review of Astronomy and Astrophysics}\ }\textbf {\bibinfo
  {volume} {47}},\ \bibinfo {pages} {291--332} (\bibinfo {year}
  {2009})}\BibitemShut {NoStop}%
\bibitem [{\citenamefont {Muñoz}(2015)}]{munoz2015}%
  \BibitemOpen
  \bibfield  {author} {\bibinfo {author} {\bibfnamefont {P.}~\bibnamefont
  {Muñoz}},\ }\emph {\bibinfo {title} {Fully kinetic PiC simulations of
  current sheet instabilities for the Solar corona}},\ \href@noop {} {Ph.D.
  thesis},\ \bibinfo  {school} {University of Göttingen} (\bibinfo {year}
  {2015})\BibitemShut {NoStop}%
\bibitem [{\citenamefont {Coppi}, \citenamefont {Greene},\ and\ \citenamefont
  {Johnson}(1966)}]{coppi1966}%
  \BibitemOpen
  \bibfield  {author} {\bibinfo {author} {\bibfnamefont {B.}~\bibnamefont
  {Coppi}}, \bibinfo {author} {\bibfnamefont {J.~M.}\ \bibnamefont {Greene}}, \
  and\ \bibinfo {author} {\bibfnamefont {J.~L.}\ \bibnamefont {Johnson}},\
  }\bibfield  {title} {\enquote {\bibinfo {title} {Resistive instabilities in a
  diffuse linear pinch},}\ }\href {\doibase 10.1088/0029-5515/6/2/003}
  {\bibfield  {journal} {\bibinfo  {journal} {Nuclear Fusion}\ }\textbf
  {\bibinfo {volume} {6}},\ \bibinfo {pages} {101} (\bibinfo {year}
  {1966})}\BibitemShut {NoStop}%
\bibitem [{\citenamefont {Pritchett}(2001)}]{pritchett2001}%
  \BibitemOpen
  \bibfield  {author} {\bibinfo {author} {\bibfnamefont {P.~L.}\ \bibnamefont
  {Pritchett}},\ }\bibfield  {title} {\enquote {\bibinfo {title} {Geospace
  environment modeling magnetic reconnection challenge: Simulations with a full
  particle electromagnetic code},}\ }\href {\doibase 10.1029/1999JA001006}
  {\bibfield  {journal} {\bibinfo  {journal} {Journal of Geophysical Research:
  Space Physics}\ }\textbf {\bibinfo {volume} {106}},\ \bibinfo {pages}
  {3783--3798} (\bibinfo {year} {2001})}\BibitemShut {NoStop}%
\bibitem [{\citenamefont {Wang}\ \emph {et~al.}(2008)\citenamefont {Wang},
  \citenamefont {Lin}, \citenamefont {Chen},\ and\ \citenamefont
  {Lin}}]{lin2008}%
  \BibitemOpen
  \bibfield  {author} {\bibinfo {author} {\bibfnamefont {X.~Y.}\ \bibnamefont
  {Wang}}, \bibinfo {author} {\bibfnamefont {Y.}~\bibnamefont {Lin}}, \bibinfo
  {author} {\bibfnamefont {L.}~\bibnamefont {Chen}}, \ and\ \bibinfo {author}
  {\bibfnamefont {Z.}~\bibnamefont {Lin}},\ }\bibfield  {title} {\enquote
  {\bibinfo {title} {A particle simulation of current sheet instabilities under
  finite guide field},}\ }\href {\doibase 10.1063/1.2938732} {\bibfield
  {journal} {\bibinfo  {journal} {Physics of Plasmas}\ }\textbf {\bibinfo
  {volume} {15}},\ \bibinfo {pages} {072103} (\bibinfo {year}
  {2008})}\BibitemShut {NoStop}%
\bibitem [{\citenamefont {Yoon}\ \emph {et~al.}(2008)\citenamefont {Yoon},
  \citenamefont {Lin}, \citenamefont {Wang},\ and\ \citenamefont
  {Lui}}]{yoon2008}%
  \BibitemOpen
  \bibfield  {author} {\bibinfo {author} {\bibfnamefont {P.~H.}\ \bibnamefont
  {Yoon}}, \bibinfo {author} {\bibfnamefont {Y.}~\bibnamefont {Lin}}, \bibinfo
  {author} {\bibfnamefont {X.~Y.}\ \bibnamefont {Wang}}, \ and\ \bibinfo
  {author} {\bibfnamefont {A.~T.~Y.}\ \bibnamefont {Lui}},\ }\bibfield  {title}
  {\enquote {\bibinfo {title} {Theory and simulation of lower-hybrid drift
  instability for current sheet with guide field},}\ }\href {\doibase
  10.1063/1.3013451} {\bibfield  {journal} {\bibinfo  {journal} {Physics of
  Plasmas}\ }\textbf {\bibinfo {volume} {15}},\ \bibinfo {pages} {112103}
  (\bibinfo {year} {2008})}\BibitemShut {NoStop}%
\bibitem [{\citenamefont {Innocenti}\ \emph {et~al.}(2016)\citenamefont
  {Innocenti}, \citenamefont {Norgen}, \citenamefont {Newman}, \citenamefont
  {Goldman}, \citenamefont {Markidis},\ and\ \citenamefont
  {Lapenta}}]{innocenti2016}%
  \BibitemOpen
  \bibfield  {author} {\bibinfo {author} {\bibfnamefont {M.~E.}\ \bibnamefont
  {Innocenti}}, \bibinfo {author} {\bibfnamefont {C.}~\bibnamefont {Norgen}},
  \bibinfo {author} {\bibfnamefont {D.}~\bibnamefont {Newman}}, \bibinfo
  {author} {\bibfnamefont {M.}~\bibnamefont {Goldman}}, \bibinfo {author}
  {\bibfnamefont {S.}~\bibnamefont {Markidis}}, \ and\ \bibinfo {author}
  {\bibfnamefont {G.}~\bibnamefont {Lapenta}},\ }\bibfield  {title} {\enquote
  {\bibinfo {title} {Study of electric and magnetic field fluctuations from
  lower hybrid drift instability waves in the terrestrial magnetotail with the
  fully kinetic, semi-implicit, adaptive multi level multi domain method},}\
  }\href {\doibase 10.1063/1.4952630} {\bibfield  {journal} {\bibinfo
  {journal} {Physics of Plasmas}\ }\textbf {\bibinfo {volume} {23}},\ \bibinfo
  {pages} {052902} (\bibinfo {year} {2016})}\BibitemShut {NoStop}%
\bibitem [{\citenamefont {Allmann-Rahn}\ \emph {et~al.}(2021)\citenamefont
  {Allmann-Rahn}, \citenamefont {Lautenbach}, \citenamefont {Grauer},\ and\
  \citenamefont {Sydora}}]{rahman2021}%
  \BibitemOpen
  \bibfield  {author} {\bibinfo {author} {\bibfnamefont {F.}~\bibnamefont
  {Allmann-Rahn}}, \bibinfo {author} {\bibfnamefont {S.}~\bibnamefont
  {Lautenbach}}, \bibinfo {author} {\bibfnamefont {R.}~\bibnamefont {Grauer}},
  \ and\ \bibinfo {author} {\bibfnamefont {R.~D.}\ \bibnamefont {Sydora}},\
  }\bibfield  {title} {\enquote {\bibinfo {title} {Fluid simulations of
  three-dimensional reconnection that capture the lower-hybrid drift
  instability},}\ }\href {\doibase 10.1017/S0022377820001683} {\bibfield
  {journal} {\bibinfo  {journal} {Journal of Plasma Physics}\ }\textbf
  {\bibinfo {volume} {87}},\ \bibinfo {pages} {905870115} (\bibinfo {year}
  {2021})}\BibitemShut {NoStop}%
\bibitem [{\citenamefont {Schmitz}\ and\ \citenamefont
  {Grauer}(2006)}]{schmitz2006}%
  \BibitemOpen
  \bibfield  {author} {\bibinfo {author} {\bibfnamefont {H.}~\bibnamefont
  {Schmitz}}\ and\ \bibinfo {author} {\bibfnamefont {R.}~\bibnamefont
  {Grauer}},\ }\bibfield  {title} {\enquote {\bibinfo {title} {Kinetic vlasov
  simulations of collisionless magnetic reconnection},}\ }\href {\doibase
  10.1063/1.2347101} {\bibfield  {journal} {\bibinfo  {journal} {Physics of
  Plasmas}\ }\textbf {\bibinfo {volume} {13}},\ \bibinfo {pages} {092309}
  (\bibinfo {year} {2006})}\BibitemShut {NoStop}%
\bibitem [{\citenamefont {Huba}, \citenamefont {Gladd},\ and\ \citenamefont
  {Papadopoulos}(1977)}]{huba1977}%
  \BibitemOpen
  \bibfield  {author} {\bibinfo {author} {\bibfnamefont {J.~D.}\ \bibnamefont
  {Huba}}, \bibinfo {author} {\bibfnamefont {N.~T.}\ \bibnamefont {Gladd}}, \
  and\ \bibinfo {author} {\bibfnamefont {K.}~\bibnamefont {Papadopoulos}},\
  }\bibfield  {title} {\enquote {\bibinfo {title} {The lower-hybrid-drift
  instability as a source of anomalous resistivity for magnetic field line
  reconnection},}\ }\href {\doibase 10.1029/GL004i003p00125} {\bibfield
  {journal} {\bibinfo  {journal} {Geophysical Research Letters}\ }\textbf
  {\bibinfo {volume} {4}},\ \bibinfo {pages} {125--128} (\bibinfo {year}
  {1977})}\BibitemShut {NoStop}%
\bibitem [{\citenamefont {Treumann}\ and\ \citenamefont
  {Baumjohann}(2013)}]{treumann2001}%
  \BibitemOpen
  \bibfield  {author} {\bibinfo {author} {\bibfnamefont {R.~A.}\ \bibnamefont
  {Treumann}}\ and\ \bibinfo {author} {\bibfnamefont {W.}~\bibnamefont
  {Baumjohann}},\ }\bibfield  {title} {\enquote {\bibinfo {title}
  {Collisionless magnetic reconnection in space plasmas},}\ }\href {\doibase
  10.3389/fphy.2013.00031} {\bibfield  {journal} {\bibinfo  {journal}
  {Frontiers in Physics}\ }\textbf {\bibinfo {volume} {1}} (\bibinfo {year}
  {2013}),\ 10.3389/fphy.2013.00031}\BibitemShut {NoStop}%
\bibitem [{\citenamefont {Graham}\ \emph {et~al.}(2019)\citenamefont {Graham},
  \citenamefont {Khotyaintsev}, \citenamefont {Vaivads},\ and\ \citenamefont
  {et~al.}}]{graham2019}%
  \BibitemOpen
  \bibfield  {author} {\bibinfo {author} {\bibfnamefont {D.~B.}\ \bibnamefont
  {Graham}}, \bibinfo {author} {\bibfnamefont {Y.~V.}\ \bibnamefont
  {Khotyaintsev}}, \bibinfo {author} {\bibfnamefont {A.}~\bibnamefont
  {Vaivads}}, \ and\ \bibinfo {author} {\bibnamefont {et~al.}},\ }\bibfield
  {title} {\enquote {\bibinfo {title} {Universality of lower hybrid waves at
  earth's magnetopause},}\ }\href {\doibase 10.1029/2019JA027155} {\bibfield
  {journal} {\bibinfo  {journal} {Journal of Geophysical Research: Space
  Physics}\ }\textbf {\bibinfo {volume} {124}},\ \bibinfo {pages} {8727--8760}
  (\bibinfo {year} {2019})}\BibitemShut {NoStop}%
\bibitem [{\citenamefont {Phan}, \citenamefont {Eastwood},\ and\ \citenamefont
  {et~al.}(2018)}]{phan2018}%
  \BibitemOpen
  \bibfield  {author} {\bibinfo {author} {\bibfnamefont {T.~D.}\ \bibnamefont
  {Phan}}, \bibinfo {author} {\bibfnamefont {J.~P.}\ \bibnamefont {Eastwood}},
  \ and\ \bibinfo {author} {\bibnamefont {et~al.}},\ }\bibfield  {title}
  {\enquote {\bibinfo {title} {Electron magnetic reconnection without ion
  coupling in earth's turbulent magnetosheath},}\ }\href {\doibase
  10.1038/s41586-018-0091-5} {\bibfield  {journal} {\bibinfo  {journal}
  {Nature}\ }\textbf {\bibinfo {volume} {557}},\ \bibinfo {pages} {202--206}
  (\bibinfo {year} {2018})}\BibitemShut {NoStop}%
\bibitem [{\citenamefont {Ergun}\ \emph {et~al.}(2017)\citenamefont {Ergun},
  \citenamefont {Chen}, \citenamefont {Wilder},\ and\ \citenamefont
  {et~al.}}]{ergun2020}%
  \BibitemOpen
  \bibfield  {author} {\bibinfo {author} {\bibfnamefont {R.~E.}\ \bibnamefont
  {Ergun}}, \bibinfo {author} {\bibfnamefont {L.-J.}\ \bibnamefont {Chen}},
  \bibinfo {author} {\bibfnamefont {V.}~\bibnamefont {Wilder}}, \ and\ \bibinfo
  {author} {\bibnamefont {et~al.}},\ }\bibfield  {title} {\enquote {\bibinfo
  {title} {Drift waves, intense parallel electric fields, and turbulence
  associated with asymmetric reconnection at the magnetopause},}\ }\href
  {\doibase 10.1002/2016GL072493} {\bibfield  {journal} {\bibinfo  {journal}
  {Geophysical Research Letters}\ }\textbf {\bibinfo {volume} {44}},\ \bibinfo
  {pages} {2978--2986} (\bibinfo {year} {2017})}\BibitemShut {NoStop}%
\bibitem [{\citenamefont {Stawarz}\ \emph {et~al.}(2016)\citenamefont
  {Stawarz}, \citenamefont {Eriksson}, \citenamefont {Wilder}, \citenamefont
  {Ergun},\ and\ \citenamefont {et~al.}}]{stawarz2021}%
  \BibitemOpen
  \bibfield  {author} {\bibinfo {author} {\bibfnamefont {J.~E.}\ \bibnamefont
  {Stawarz}}, \bibinfo {author} {\bibfnamefont {S.}~\bibnamefont {Eriksson}},
  \bibinfo {author} {\bibfnamefont {F.~D.}\ \bibnamefont {Wilder}}, \bibinfo
  {author} {\bibfnamefont {R.~E.}\ \bibnamefont {Ergun}}, \ and\ \bibinfo
  {author} {\bibnamefont {et~al.}},\ }\bibfield  {title} {\enquote {\bibinfo
  {title} {Observations of turbulence in a kelvin-helmholtz event on 8
  september 2015 by the magnetospheric multiscale mission},}\ }\href {\doibase
  10.1002/2016JA023458} {\bibfield  {journal} {\bibinfo  {journal} {Journal of
  Geophysical Research: Space Physics}\ }\textbf {\bibinfo {volume} {121}},\
  \bibinfo {pages} {11,021--11,034} (\bibinfo {year} {2016})}\BibitemShut
  {NoStop}%
\bibitem [{\citenamefont {Gary}(1993)}]{gary1993}%
  \BibitemOpen
  \bibfield  {author} {\bibinfo {author} {\bibfnamefont {S.~P.}\ \bibnamefont
  {Gary}},\ }\href {\doibase 10.1017/CBO9780511551512} {\emph {\bibinfo {title}
  {Theory of Space Plasma Microinstabilities}}}\ (\bibinfo  {publisher}
  {Cambridge University Press},\ \bibinfo {year} {1993})\BibitemShut {NoStop}%
\bibitem [{\citenamefont {Ng}\ \emph {et~al.}(2023)\citenamefont {Ng},
  \citenamefont {Yoo}, \citenamefont {Chen}, \citenamefont {Bessho},\ and\
  \citenamefont {Ji}}]{ng2023}%
  \BibitemOpen
  \bibfield  {author} {\bibinfo {author} {\bibfnamefont {J.}~\bibnamefont
  {Ng}}, \bibinfo {author} {\bibfnamefont {J.}~\bibnamefont {Yoo}}, \bibinfo
  {author} {\bibfnamefont {L.-J.}\ \bibnamefont {Chen}}, \bibinfo {author}
  {\bibfnamefont {N.}~\bibnamefont {Bessho}}, \ and\ \bibinfo {author}
  {\bibfnamefont {H.}~\bibnamefont {Ji}},\ }\bibfield  {title} {\enquote
  {\bibinfo {title} {3d simulation of lower-hybrid drift waves in strong guide
  field asymmetric reconnection in laboratory experiments},}\ }\href {\doibase
  10.1063/5.013827} {\bibfield  {journal} {\bibinfo  {journal} {physics of
  plasmas}\ }\textbf {\bibinfo {volume} {30}},\ \bibinfo {pages} {042101}
  (\bibinfo {year} {2023})}\BibitemShut {NoStop}%
\bibitem [{\citenamefont {Muñoz}\ \emph {et~al.}(2015)\citenamefont {Muñoz},
  \citenamefont {Told}, \citenamefont {Kilian}, \citenamefont {Büchner},\ and\
  \citenamefont {Jenko}}]{told2011}%
  \BibitemOpen
  \bibfield  {author} {\bibinfo {author} {\bibfnamefont {P.}~\bibnamefont
  {Muñoz}}, \bibinfo {author} {\bibfnamefont {D.}~\bibnamefont {Told}},
  \bibinfo {author} {\bibfnamefont {P.}~\bibnamefont {Kilian}}, \bibinfo
  {author} {\bibfnamefont {J.}~\bibnamefont {Büchner}}, \ and\ \bibinfo
  {author} {\bibfnamefont {F.}~\bibnamefont {Jenko}},\ }\bibfield  {title}
  {\enquote {\bibinfo {title} {Gyrokinetic and kinetic particle-in-cell
  simulations of guide-field reconnection. i: Macroscopic effects of the
  electron flows},}\ }\href {\doibase 10.1063/1.4928381} {\bibfield  {journal}
  {\bibinfo  {journal} {Physics of Plasmas}\ }\textbf {\bibinfo {volume}
  {22}},\ \bibinfo {pages} {082110} (\bibinfo {year} {2015})}\BibitemShut
  {NoStop}%
\bibitem [{\citenamefont {Pueschel}\ \emph {et~al.}(2011)\citenamefont
  {Pueschel}, \citenamefont {Told}, \citenamefont {Büchner},\ and\
  \citenamefont {Jenko}}]{pueschel2011}%
  \BibitemOpen
  \bibfield  {author} {\bibinfo {author} {\bibfnamefont {M.~J.}\ \bibnamefont
  {Pueschel}}, \bibinfo {author} {\bibfnamefont {D.}~\bibnamefont {Told}},
  \bibinfo {author} {\bibfnamefont {J.}~\bibnamefont {Büchner}}, \ and\
  \bibinfo {author} {\bibfnamefont {F.}~\bibnamefont {Jenko}},\ }\bibfield
  {title} {\enquote {\bibinfo {title} {Gyrokinetic simulation of magnetic
  reconnection},}\ }\href {\doibase 10.1063/1.3656965} {\bibfield  {journal}
  {\bibinfo  {journal} {Physics of Plasmas}\ }\textbf {\bibinfo {volume}
  {18}},\ \bibinfo {pages} {112102} (\bibinfo {year} {2011})}\BibitemShut
  {NoStop}%
\bibitem [{\citenamefont {Thatikonda}()}]{your2025paper}%
  \BibitemOpen
  \bibfield  {author} {\bibinfo {author} {\bibfnamefont {S.~c.}\ \bibnamefont
  {Thatikonda}},\ }\href@noop {} {\enquote {\bibinfo {title} {Verification of a
  hybrid gyrokinetic model using the advanced semi-lagrange code ssv},}\
  }\bibinfo {note} {Submitted to Computer Physics Communications,
  2025}\BibitemShut {NoStop}%
\bibitem [{\citenamefont {Ricci}\ \emph {et~al.}(2005)\citenamefont {Ricci},
  \citenamefont {Brackbill}, \citenamefont {Daughton},\ and\ \citenamefont
  {Lapenta}}]{Ricci2005}%
  \BibitemOpen
  \bibfield  {author} {\bibinfo {author} {\bibfnamefont {P.}~\bibnamefont
  {Ricci}}, \bibinfo {author} {\bibfnamefont {J.~U.}\ \bibnamefont
  {Brackbill}}, \bibinfo {author} {\bibfnamefont {W.}~\bibnamefont {Daughton}},
  \ and\ \bibinfo {author} {\bibfnamefont {G.}~\bibnamefont {Lapenta}},\
  }\bibfield  {title} {\enquote {\bibinfo {title} {New role of the lower-hybrid
  drift instability in magnetic reconnection},}\ }\href {\doibase
  10.1063/1.1885002} {\bibfield  {journal} {\bibinfo  {journal} {Physics of
  Plasmas}\ }\textbf {\bibinfo {volume} {12}},\ \bibinfo {pages} {055901}
  (\bibinfo {year} {2005})}\BibitemShut {NoStop}%
\bibitem [{\citenamefont {Mandell}\ \emph {et~al.}(2020)\citenamefont
  {Mandell}, \citenamefont {Hakim}, \citenamefont {Hammett}, \citenamefont
  {Stoltzfus-Dueck},\ and\ \citenamefont {Belli}}]{Mandell2020}%
  \BibitemOpen
  \bibfield  {author} {\bibinfo {author} {\bibfnamefont {N.~R.}\ \bibnamefont
  {Mandell}}, \bibinfo {author} {\bibfnamefont {A.}~\bibnamefont {Hakim}},
  \bibinfo {author} {\bibfnamefont {G.~W.}\ \bibnamefont {Hammett}}, \bibinfo
  {author} {\bibfnamefont {T.}~\bibnamefont {Stoltzfus-Dueck}}, \ and\ \bibinfo
  {author} {\bibfnamefont {E.}~\bibnamefont {Belli}},\ }\bibfield  {title}
  {\enquote {\bibinfo {title} {Electromagnetic full-$f$ gyrokinetics in the
  tokamak edge with discontinuous galerkin methods.}}\ }\href {\doibase
  10.1017/S0022377820000070} {\bibfield  {journal} {\bibinfo  {journal}
  {Journal of Plasma Physics}\ }\textbf {\bibinfo {volume} {86}} (\bibinfo
  {year} {2020}),\ 10.1017/S0022377820000070}\BibitemShut {NoStop}%
\bibitem [{\citenamefont {Pezzi}\ \emph {et~al.}(2019)\citenamefont {Pezzi},
  \citenamefont {Valentini}, \citenamefont {Perrone}, \citenamefont {Servidio},
  \citenamefont {Veltri}, \citenamefont {Malara}, \citenamefont {Tenerani},
  \citenamefont {Franci},\ and\ \citenamefont {Matthaeus}}]{Pezzi2019}%
  \BibitemOpen
  \bibfield  {author} {\bibinfo {author} {\bibfnamefont {O.}~\bibnamefont
  {Pezzi}}, \bibinfo {author} {\bibfnamefont {F.}~\bibnamefont {Valentini}},
  \bibinfo {author} {\bibfnamefont {D.}~\bibnamefont {Perrone}}, \bibinfo
  {author} {\bibfnamefont {S.}~\bibnamefont {Servidio}}, \bibinfo {author}
  {\bibfnamefont {P.}~\bibnamefont {Veltri}}, \bibinfo {author} {\bibfnamefont
  {F.}~\bibnamefont {Malara}}, \bibinfo {author} {\bibfnamefont
  {A.}~\bibnamefont {Tenerani}}, \bibinfo {author} {\bibfnamefont
  {L.}~\bibnamefont {Franci}}, \ and\ \bibinfo {author} {\bibfnamefont {W.~H.}\
  \bibnamefont {Matthaeus}},\ }\bibfield  {title} {\enquote {\bibinfo {title}
  {Vida: a vlasov--darwin solver for plasma physics at electron scales.}}\
  }\href {\doibase 10.1017/S0022377819000631} {\bibfield  {journal} {\bibinfo
  {journal} {Journal of Plasma Physics}\ }\textbf {\bibinfo {volume} {85}},\
  \bibinfo {pages} {905850506} (\bibinfo {year} {2019})}\BibitemShut {NoStop}%
\bibitem [{\citenamefont {Dannert}\ and\ \citenamefont
  {Jenko}(2004)}]{Dannert2004}%
  \BibitemOpen
  \bibfield  {author} {\bibinfo {author} {\bibfnamefont {T.}~\bibnamefont
  {Dannert}}\ and\ \bibinfo {author} {\bibfnamefont {F.}~\bibnamefont
  {Jenko}},\ }\bibfield  {title} {\enquote {\bibinfo {title} {Vlasov simulation
  of kinetic shear alfven waves.}}\ }\href {\doibase 10.1016/j.cpc.2004.09.001}
  {\bibfield  {journal} {\bibinfo  {journal} {Computer Physics Communications}\
  }\textbf {\bibinfo {volume} {163}},\ \bibinfo {pages} {67--78} (\bibinfo
  {year} {2004})}\BibitemShut {NoStop}%
\bibitem [{\citenamefont {Tanaka}(2017)}]{Tanaka2017}%
  \BibitemOpen
  \bibfield  {author} {\bibinfo {author} {\bibfnamefont {S.}~\bibnamefont
  {Tanaka}},\ }\bibfield  {title} {\enquote {\bibinfo {title} {Multidimensional
  vlasov--poisson simulations with high-order monotonicity- and
  positivity-preserving schemes.}}\ }\href {\doibase 10.3847/1538-4357/aa901f}
  {\bibfield  {journal} {\bibinfo  {journal} {The Astrophysical Journal}\
  }\textbf {\bibinfo {volume} {849}},\ \bibinfo {pages} {76} (\bibinfo {year}
  {2017})}\BibitemShut {NoStop}%
\bibitem [{\citenamefont {Cranmer}(2002)}]{Cranmer2002}%
  \BibitemOpen
  \bibfield  {author} {\bibinfo {author} {\bibfnamefont {S.}~\bibnamefont
  {Cranmer}},\ }\bibfield  {title} {\enquote {\bibinfo {title} {Coronal holes
  and the solar wind},}\ }in\ \href@noop {} {\emph {\bibinfo {booktitle}
  {Multi-wavelength Observations of Coronal Structure and Dynamics}}},\
  \bibinfo {series} {COSPAR Colloquia Series}, Vol.~\bibinfo {volume} {13},\
  \bibinfo {editor} {edited by\ \bibinfo {editor} {\bibfnamefont {P.~C.}\
  \bibnamefont {Martens}}\ and\ \bibinfo {editor} {\bibfnamefont {D.~P.}\
  \bibnamefont {Cauffman}}}\ (\bibinfo  {publisher} {Pergamon},\ \bibinfo
  {year} {2002})\ pp.\ \bibinfo {pages} {3--12}\BibitemShut {NoStop}%
\bibitem [{\citenamefont {Oughton}\ \emph {et~al.}(2011)\citenamefont
  {Oughton}, \citenamefont {Matthaeus}, \citenamefont {Smith}, \citenamefont
  {Breech},\ and\ \citenamefont {Isenberg}}]{Oughton2011}%
  \BibitemOpen
  \bibfield  {author} {\bibinfo {author} {\bibfnamefont {S.}~\bibnamefont
  {Oughton}}, \bibinfo {author} {\bibfnamefont {W.~H.}\ \bibnamefont
  {Matthaeus}}, \bibinfo {author} {\bibfnamefont {C.~W.}\ \bibnamefont
  {Smith}}, \bibinfo {author} {\bibfnamefont {B.}~\bibnamefont {Breech}}, \
  and\ \bibinfo {author} {\bibfnamefont {P.~A.}\ \bibnamefont {Isenberg}},\
  }\bibfield  {title} {\enquote {\bibinfo {title} {Transport of solar wind
  fluctuations: A two-component model},}\ }\href {\doibase
  10.1029/2010JA016365} {\bibfield  {journal} {\bibinfo  {journal} {Journal of
  Geophysical Research: Space Physics}\ }\textbf {\bibinfo {volume} {116}}
  (\bibinfo {year} {2011}),\ 10.1029/2010JA016365}\BibitemShut {NoStop}%
\bibitem [{\citenamefont {Podesta}(2012)}]{Podesta2012}%
  \BibitemOpen
  \bibfield  {author} {\bibinfo {author} {\bibfnamefont {J.~J.}\ \bibnamefont
  {Podesta}},\ }\bibfield  {title} {\enquote {\bibinfo {title} {The need to
  consider ion bernstein waves as a dissipation channel of solar wind
  turbulence.}}\ }\href {\doibase 10.1029/2012JA017770} {\bibfield  {journal}
  {\bibinfo  {journal} {Journal of Geophysical Research: Space Physics}\
  }\textbf {\bibinfo {volume} {117}} (\bibinfo {year} {2012}),\
  10.1029/2012JA017770}\BibitemShut {NoStop}%
\bibitem [{\citenamefont {Howes}\ \emph {et~al.}(2011)\citenamefont {Howes},
  \citenamefont {TenBarge}, \citenamefont {Dorland}, \citenamefont {Quataert},
  \citenamefont {Schekochihin}, \citenamefont {Numata},\ and\ \citenamefont
  {Tatsuno}}]{Howes2011}%
  \BibitemOpen
  \bibfield  {author} {\bibinfo {author} {\bibfnamefont {G.~G.}\ \bibnamefont
  {Howes}}, \bibinfo {author} {\bibfnamefont {J.~M.}\ \bibnamefont {TenBarge}},
  \bibinfo {author} {\bibfnamefont {W.}~\bibnamefont {Dorland}}, \bibinfo
  {author} {\bibfnamefont {E.}~\bibnamefont {Quataert}}, \bibinfo {author}
  {\bibfnamefont {A.~A.}\ \bibnamefont {Schekochihin}}, \bibinfo {author}
  {\bibfnamefont {R.}~\bibnamefont {Numata}}, \ and\ \bibinfo {author}
  {\bibfnamefont {T.}~\bibnamefont {Tatsuno}},\ }\bibfield  {title} {\enquote
  {\bibinfo {title} {Gyrokinetic simulations of solar wind turbulence from ion
  to electron scales.}}\ }\href {\doibase 10.1103/PhysRevLett.107.035004}
  {\bibfield  {journal} {\bibinfo  {journal} {Physical Review Letters}\
  }\textbf {\bibinfo {volume} {107}},\ \bibinfo {pages} {035004} (\bibinfo
  {year} {2011})}\BibitemShut {NoStop}%
\bibitem [{\citenamefont {Chen}(2016)}]{Chen2016}%
  \BibitemOpen
  \bibfield  {author} {\bibinfo {author} {\bibfnamefont {C.~H.~K.}\
  \bibnamefont {Chen}},\ }\bibfield  {title} {\enquote {\bibinfo {title}
  {Recent progress in astrophysical plasma turbulence from solar wind
  observations.}}\ }\href {\doibase 10.1017/S0022377816001124} {\bibfield
  {journal} {\bibinfo  {journal} {Journal of Plasma Physics}\ }\textbf
  {\bibinfo {volume} {82}} (\bibinfo {year} {2016}),\
  10.1017/S0022377816001124}\BibitemShut {NoStop}%
\bibitem [{\citenamefont {Hollweg}(1978)}]{Hollweg1978}%
  \BibitemOpen
  \bibfield  {author} {\bibinfo {author} {\bibfnamefont {J.~V.}\ \bibnamefont
  {Hollweg}},\ }\bibfield  {title} {\enquote {\bibinfo {title} {A quasi-linear
  wkb kinetic theory for nonplanar waves in a nonhomogeneous warm plasma, 1.
  transverse waves propagating along axisymmetric $b_0$},}\ }\href {\doibase
  10.1029/JA083iA02p00563} {\bibfield  {journal} {\bibinfo  {journal} {Journal
  of Geophysical Research: Space Physics}\ }\textbf {\bibinfo {volume} {83}},\
  \bibinfo {pages} {563--574} (\bibinfo {year} {1978})}\BibitemShut {NoStop}%
\bibitem [{\citenamefont {Karimabadi}\ \emph {et~al.}(2014)\citenamefont
  {Karimabadi}, \citenamefont {Roytershteyn}, \citenamefont {Vu}, \citenamefont
  {Omelchenko}, \citenamefont {Scudder}, \citenamefont {Daughton},
  \citenamefont {Dimmock}, \citenamefont {Nykyri}, \citenamefont {Wan},
  \citenamefont {Sibeck}, \citenamefont {Tatineni}, \citenamefont {Majumdar},
  \citenamefont {Loring},\ and\ \citenamefont {Geveci}}]{Karimabadi2014}%
  \BibitemOpen
  \bibfield  {author} {\bibinfo {author} {\bibfnamefont {H.}~\bibnamefont
  {Karimabadi}}, \bibinfo {author} {\bibfnamefont {V.}~\bibnamefont
  {Roytershteyn}}, \bibinfo {author} {\bibfnamefont {H.~X.}\ \bibnamefont
  {Vu}}, \bibinfo {author} {\bibfnamefont {Y.~A.}\ \bibnamefont {Omelchenko}},
  \bibinfo {author} {\bibfnamefont {J.}~\bibnamefont {Scudder}}, \bibinfo
  {author} {\bibfnamefont {W.}~\bibnamefont {Daughton}}, \bibinfo {author}
  {\bibfnamefont {A.}~\bibnamefont {Dimmock}}, \bibinfo {author} {\bibfnamefont
  {K.}~\bibnamefont {Nykyri}}, \bibinfo {author} {\bibfnamefont
  {M.}~\bibnamefont {Wan}}, \bibinfo {author} {\bibfnamefont {D.}~\bibnamefont
  {Sibeck}}, \bibinfo {author} {\bibfnamefont {M.}~\bibnamefont {Tatineni}},
  \bibinfo {author} {\bibfnamefont {A.}~\bibnamefont {Majumdar}}, \bibinfo
  {author} {\bibfnamefont {B.}~\bibnamefont {Loring}}, \ and\ \bibinfo {author}
  {\bibfnamefont {B.}~\bibnamefont {Geveci}},\ }\bibfield  {title} {\enquote
  {\bibinfo {title} {The link between shocks, turbulence, and magnetic
  reconnection in collisionless plasmas},}\ }\href {\doibase 10.1063/1.4882875}
  {\bibfield  {journal} {\bibinfo  {journal} {Physics of Plasmas}\ }\textbf
  {\bibinfo {volume} {21}},\ \bibinfo {pages} {062308} (\bibinfo {year}
  {2014})}\BibitemShut {NoStop}%
\bibitem [{\citenamefont {deOliveira Lopes}(2022)}]{Lopes2022}%
  \BibitemOpen
  \bibfield  {author} {\bibinfo {author} {\bibfnamefont {F.~N.}\ \bibnamefont
  {deOliveira Lopes}},\ }\emph {\bibinfo {title} {Geometrical Formulation of
  Hybrid Kinetic and Gyrokinetic Hamiltonian Field Theory for Astrophysical and
  Laboratory Plasmas.}},\ \href {https://doi.org/10.13154/294-9291} {\bibinfo
  {type} {Ph.d. thesis}},\ \bibinfo  {school} {Ruhr-Universität Bochum}
  (\bibinfo {year} {2022})\BibitemShut {NoStop}%
\end{thebibliography}%

\end{document}